\documentclass[sigconf,authorversion=true]{acmart}
\AtBeginDocument{%
  }

\copyrightyear{2025}
\acmYear{2025}
\setcopyright{acmlicensed}\acmConference[CHI '25]{CHI Conference on Human Factors in Computing Systems}{April 26-May 1, 2025}{Yokohama, Japan}
\acmBooktitle{CHI Conference on Human Factors in Computing Systems (CHI '25), April 26-May 1, 2025, Yokohama, Japan}
\acmDOI{10.1145/3706598.3713346}
\acmISBN{979-8-4007-1394-1/25/04}

\usepackage{array}
\usepackage{url}
\usepackage{xcolor}
\usepackage{multirow}
\usepackage{ifthen}

\usepackage{subcaption}

\newboolean{showrevisions}
\setboolean{showrevisions}{false}

\ifthenelse{\boolean{showrevisions}}{\definecolor{newcolor}{rgb}{0.15, 0.15, 1}
\definecolor{chcolor}{rgb}{0.15, 0.5, 0.15}}{\definecolor{newcolor}{rgb}{0, 0, 0}
\definecolor{chcolor}{rgb}{0, 0, 0}}

\newcommand{\newtext}[1]{\textcolor{newcolor}{#1}}

\newcommand{\changetext}[1]{\textcolor{chcolor}{#1}}

\begin{document}

\title{Examining the Effects of Immersive and Non-Immersive Presenter Modalities on Engagement and Social Interaction in Co-located Augmented Presentations}
\renewcommand{\shorttitle}{Effects of Immersive and Non-Immersive Presenter Modalities in Co-located Augmented Presentations}
% Examining the Effects of Immersive and Non-Immersive Presenter Modalities on Engagement and Social Interaction in Co-located Augmented Presentations

% Examining Symetric and Asymetric Techinques for Co-Located Immersive Presentations
% and their effects on engagement / social interaction
% 
% 

% Exploring Social Interaction and Engagement in a Co-located Immersive Presentation System

% Immersive presentation
% Augmented presentaion
% Engaging
% Social interaction
% Co-located

% \affiliation{%
\author{Matt Gottsacker}
\affiliation{%
% \department{Global Technology Applied Research}
  \institution{JPMorganChase}
  \city{New York}
  \state{NY}
  \country{USA}
}
\affiliation{%
% \department{SREAL}
  \institution{University of Central Florida}
  \city{Orlando}
  \state{FL}
  \country{USA}
}
% \authornote{}
\email{matt.gottsacker@gmail.com}
\orcid{0000-0002-3575-1133}

\author{Mengyu Chen}
\affiliation{%
% \department{Global Technology Applied Research}
  \institution{JPMorganChase}
  \city{New York}
  \state{NY}
  \country{USA}
}
\email{mengyu.chen@jpmchase.com}
\orcid{0000-0001-6833-7273}

\author{David Saffo}
\affiliation{%
% \department{Global Technology Applied Research}
  \institution{JPMorganChase}
  \city{New York}
  \state{NY}
  \country{USA}
}
\email{david.saffo@jpmchase.com}
\orcid{0000-0001-9515-048X}

\author{Feiyu Lu}
\affiliation{%
% \department{Global Technology Applied Research}
  \institution{JPMorganChase}
  \city{New York}
  \state{NY}
  \country{USA}
}
\email{feiyu.lu@jpmchase.com}
\orcid{0000-0002-1939-9352}

\author{Benjamin Lee}
\affiliation{%
% \department{Global Technology Applied Research}
  \institution{JPMorganChase}
  \city{New York}
  \state{NY}
  \country{USA}
}
\email{benjamin.lee@jpmchase.com}
\orcid{0000-0002-1171-4741}

\author{Blair MacIntyre}
\affiliation{%
% \department{Global Technology Applied Research}
  \institution{JPMorganChase}
  \city{New York}
  \state{NY}
  \country{USA}
}
\email{blair.macintyre@jpmchase.com}
\orcid{0000-0002-5357-2366}
\renewcommand{\shortauthors}{Gottsacker et al.}

\begin{abstract}
Head-worn augmented reality (AR) allows audiences to be immersed and engaged in stories told by live presenters. 
While presenters may also be in AR to have the same level of immersion and awareness as their audience, this symmetric presentation style may diminish important social cues such as eye contact.
In this work, we examine the effects this (a)symmetry has on engagement, group awareness, and social interaction in co-located one-on-one augmented presentations.
We developed a presentation system incorporating 2D/3D content that audiences can view and interact with in AR, with presenters controlling and delivering the presentation in either a symmetric style in AR, or an asymmetric style with a handheld tablet. 
We conducted a within- and between-subjects evaluation with 12 participant pairs to examine the differences between these symmetric and asymmetric presentation modalities.
From our findings, we extracted four themes and derived strategies and guidelines for designers interested in augmented presentations.   
\end{abstract}

\begin{CCSXML}
<ccs2012>
   <concept>
       <concept_desc>Human-centered computing~Mixed / augmented reality</concept_desc>
       </concept>
   <concept>
       <concept_desc>Human-centered computing~Collaborative interaction</concept_desc>
       </concept>
   <concept>
       <concept_desc>Human-centered computing~User studies</concept_desc>
       </concept>
 </ccs2012>
\end{CCSXML}

\ccsdesc[500]{Human-centered computing~Mixed / augmented reality}
\ccsdesc[500]{Human-centered computing~Collaborative interaction}
\ccsdesc[500]{Human-centered computing~User studies}

\keywords{Augmented Reality, Augmented Presentation, Social Interaction, Engagement}
\begin{teaserfigure}
    \centering
  \includegraphics[width=0.74\textwidth]{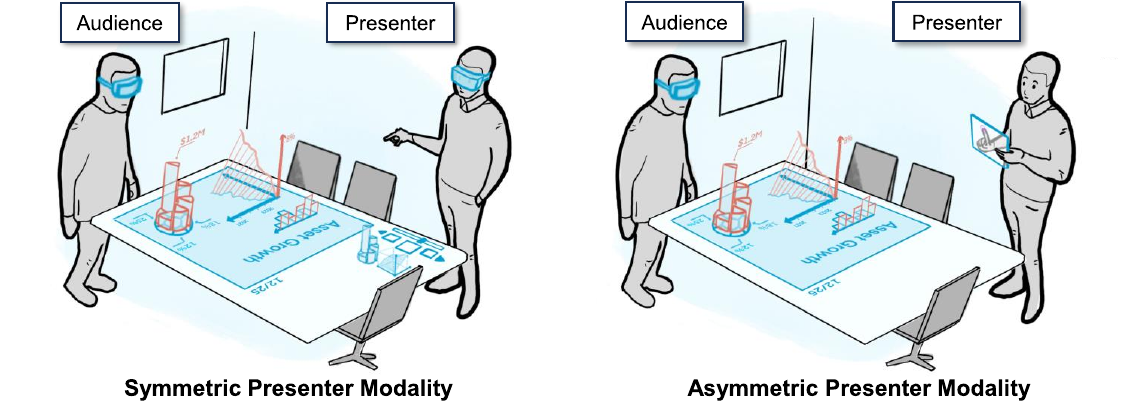}
  \caption{Illustrations of the immersive presentation system with symmetric and asymmetric presentation modalities.
  In the symmetric modality, the presenter uses a head-worn augmented reality (AR) display to control the presentation.
  In the asymmetric modality, the presenter controls the presentation with a tablet.}
  \Description{Illustrations of the immersive presentation system.
  A presenter and audience member are positioned around a real-world table with virtual content overlaid on top of it.
  The audience is wearing an augmented reality head-worn display to view the virtual content.
  In the left image, the presenter uses a tablet to control the presentation.
  In the right image, the presenter is also wearing an augmented reality head-worn display to control the presentation.
  }
  \label{fig:teaser}
\end{teaserfigure}

% \received{12 September 2024}

\maketitle

\section{Introduction}

One of the greatest strengths of immersive technologies, such as augmented and virtual reality (AR/VR), is its ability to facilitate engaging and captivating experiences. 
Researchers, designers, and creative technologists have explored this strength for applications in education~\cite{kuhail2022}, healthcare~\cite{Ryan2022}, conferencing~\cite{Moreira2023}, entertainment~\cite{Donghee2019}, marketing~\cite{SCHOLZ2016149}, and more.
Across these use cases, achieving higher levels of user engagement increases outcomes for learning, retention, productivity, relations, collaboration, and overall satisfaction.  
Among the engagement-focused immersive experience use cases, augmented and immersive presentations have emerged as topics of interest. 
Presentations --- the delivery of information, concepts, or arguments --- come in various forms and contexts, each tailored to specific purposes and audiences. 
Formal presentations use structured content and visual aids, while informal presentations are more casual and conversational, and interactive presentations involve audience participation. 
Regardless of the format, effective presentations require audience engagement, which can be enhanced through visual aids, storytelling, interactive elements, and real-time feedback mechanisms.
The premise of \textit{augmented presentation} is to leverage the affordances of AR/VR, such as spatial presentation, embodied interaction, and remote collaboration, to enable engaging, informative, and memorable presentations. 

% Introducing higher levels of immersion to both audiences and presenters alike requires new considerations, however. 
% %Social interaction and group awareness are co-factors in the overall engagement and outcomes of presentations. 
% Behaviors such as eye contact, body language cues, and verbal and non-verbal communication are all social interactions presenters and audiences use to connect interpersonally. 
% Similarly, presenters and audiences rely on group awareness (i.e., understanding what each other are attending to and referring to)
% % inter-referential awareness, common ground, and group awareness 
% to ensure effective communication, interaction, and understanding of the presentation topic and materials. 
% In augmented presentation, head-worn devices (HWD) are used to enable spatial presentation and interaction. 
% While these devices enable higher levels of immersion in the virtual world, they can diminish the level of presence in the physical world, making co-located social interaction and group awareness more difficult. 
% To combat this, the level of immersion for presenters and audiences can be varied by using different devices with affordances better suited for the respective roles.  

In co-located one-on-one presentation contexts (e.g., financial planning meetings, medical consultations, or creative design reviews) a single presenter controls the presentation flow for a single audience member who consumes and inquires about the presented information. 
% During an augmented presentation where the presenter and audience share the same level of immersion using symmetric HWD presentation modalities, participants share equal affordances for viewing and controlling the content in the virtual world.
In an augmented presentation setting, the presenter and audience may both use AR head-worn displays (HWDs) to have symmetric affordances for viewing and controlling the virtual presentation content.
This symmetry could better enable presenters and audiences to have high group awareness (i.e., understanding what each other are attending to and referring to) to ensure effective communication, interaction, and understanding of the presentation topic and materials.
% Behaviors such as eye contact, body language cues, and verbal and non-verbal communication are all social interactions presenters and audiences use to connect interpersonally. 
However, even with AR HWDs, important social interaction cues such as eye contact could be diminished due to the wearer's face being obstructed~\cite{mcatamney2006examination}, which can
% reduce cues that people use to 
impair how people
regulate conversation flow, express and understand emotions, and direct attention~\cite{bassili1979emotion, garau2001gaze}.
% Solutions that render the HWD user's eyes on the outside of the display, such as in the Apple Vision Pro\footnote{\url{https://www.apple.com/apple-vision-pro/}} and research prototypes~\cite{bozgeyikli2024googly, chan2017frontface, mai2017transparent, matsuda2021reverse} currently have limitations in terms of their rendering and display quality.
Additionally, presenters may not want to wear an HWD for extended periods due to discomfort and fatigue~\cite{biener2022quantifying}, especially if
they need to conduct multiple presentations back-to-back.
% multiple presentations were to be conducted back-to-back.
In the early stages of this work, we observed this reluctance 
% for presenters to use an AR HWD 
from stakeholders at our organization
% , of whom 
who regularly lead co-located one-on-one presentations. They emphasized the importance of social interaction in these presentations, which would likely be obstructed should they use an HWD.
% In early design work for this project, we interviewed several stakeholders at our organization who regularly lead co-located one-on-one presentations.
% They emphasized the importance of unobstructed social interaction in meetings and expressed a desire to avoid having the presenter use an HWD.
For these reasons, we explored asymmetric presenter modalities, such as a non-immersive tablet interface, to trade a presenter's level of immersion and group awareness in the symmetric presentation modality
% spatial referencing common ground 
for possibly higher levels of social interaction.
% Instead, we can introduce asymmetric presentation modalities, such as a tablet interface, to trade a presenter's level of immersion and spatial referencing common ground for possibly higher levels of social interaction.
\newtext{
However, we chose to always have the audience view the augmented presentation with an AR HWD to prioritize their immersion in the experience.
}

\changetext{
In this paper, we present an exploratory user study with (n = 12) dyads examining how symmetric and asymmetric presenter modalities affect participants' presentation experiences regarding their interpersonal engagement and group awareness during the presentation.
To conduct this study, we created a prototype presentation system with which an audience member with an AR HWD can view virtual presentation content, whilst a presenter can view and control the same presentation content using either a symmetric interface (i.e., also with an AR HWD) or an asymmetric interface (i.e., with a tablet).
As asymmetry can result in reducing the common ground between users, our asymmetric interface provides additional features designed to maintain group awareness between the presenter and audience, including simulated mirrored audience views and the ability for the presenter to point to virtual content through those views.}
\changetext{We conducted semi-structured interviews with participants about their engagement, group awareness, and overall experiences.
Additionally,} to study interpersonal engagement, we observed participants' social behaviors (e.g., how participants express and share social cues such as gestures and eye contact).
For group awareness, we observed participants' referencing behaviors (e.g., how participants referred to presentation content spatially when directing each other's attention).
Last, we collected questionnaire responses about presenters' cognitive load and presenters' and audience members' perceptions of engagement with the presentation (e.g., what was engaging) and their group awareness (e.g., how they referred to content or assessed their partner's attention) when using each interface.
% We also conducted semi-structured interviews with participants about their engagement and group awareness.
Our results suggest the benefits of immersive augmented presentation systems, strengths for each presenter modality, and open challenges for providing engaging presentation experiences.

\changetext{The contributions of this work are as follows:}
\changetext{\begin{itemize}
    \item An exploratory user study that examines the effects of symmetric and asymmetric presenter modalities on participants' social behaviors and group awareness.
    \item Four themes, derived from analyzing participants' user experience, around presentation techniques, spatial reference, social interaction, and interface impact.
    \item Design guidelines for researchers and practitioners interested in creating effective immersive presentation systems, with insights about trade-offs for selecting between symmetric and asymmetric presentation formats in co-located settings.
    \item The design of a novel prototype system for interactive, immersive presentations between physically co-located users including both 2D and 3D content and with symmetric and asymmetric presenter control modalities and interfaces.
    % on presentation format selection and content preparation considerations.
\end{itemize}
}
\section{Background}
Our work is inspired by and grounded in three key areas of research: immersive storytelling and presentation, symmetric and asymmetric mixed-reality systems, and social interactions in cross-reality environments.

\subsection{Immersive Storytelling and Presentation}
Presenters often rely on captivating visual imagery to draw in audiences and increase their emotional and cognitive engagement with the narrative~\cite{Lackmann2021TheIO, WANG2020106279, KING2020104797}. The desire to further bolster this engagement had given rise to a form of storytelling that closely blends the visual elements with the presenter's verbal statements and body movements, thus giving the illusion that the presenter is physically ``in control'' and in sync with the displayed information. Ever since Hans Rosling's premiere example of this presentation style~\cite{rosling2010}, research has strived towards making this form of storytelling more accessible and usable for laypeople. Said works have explored how interaction styles such as sketching~\cite{perlin2018chalktalk}, body gestures~\cite{hall2022chironomia, saquib2019}, and speech~\cite{liao2022} can further integrate the presenter's actions with visual elements in real-time.
Yet, this form of storytelling treats the audience simply as a passive observer with little to no agency within the presented narrative. While acceptable for presentations to the masses, such as webinars, it is likely inadequate for more intimate co-located presentations where the audience's input and engagement are paramount (e.g., in sales pitches), especially when dialogue occurs between them and the presenter.

\newtext{
Prior research has demonstrated benefits of interactive elements in traditional 2D presentations in educational settings.
For example, researchers have found that real-time audience polling systems for multiple-choice questions foster active participation and improve knowledge retention among medical professionals and patients~\cite{grzeskowiak2015enhancing, stoevesandt2020interactive}.
Additionally, systems that allow students to interact with presentation content more directly using their devices (e.g., by sketching or writing open-ended responses on their versions of the slides) enhance STEM students' learning and engagement~\cite{shatri2022interactive, haryani2021impact, ruado2024enhancing}.
Therefore, we sought to provide interactivity for the audience to help drive their engagement in the presentation.
}

\newtext{
Recent work has explored using interactive storytelling in AR contexts as well.
Inspired in part by immersive theater performances involving spatial audience attention direction and multimodal interaction~\cite{white2012theatre}, 
Zhang et al.~\cite{zhang2024jigsaw} introduced a system for creating and experiencing interactive immersive narratives using AR and IoT technology, which audience members found memorable and engaging.
}
Additionally, Gottsacker et al.~\cite{gottsacker2023presentation} proposed using head-worn AR to involve the audience in \changetext{an augmented} presentation directly. In it, the audience sees and can interact with visual elements in AR while the presenter narrates and controls the presentation via a tablet. While this approach encourages the audience to engage with the content for themselves directly, it still causes a visual separation \changetext{for the audience} between the \changetext{input and output of the presenter's interactions}, 
% content and presenter, 
which the aforementioned presentation styles~\cite{hall2022chironomia, perlin2018chalktalk, saquib2019, liao2022} had sought to remove. They also did not evaluate the usability or effectiveness of the new presentation format.

Our work, therefore, seeks to compare the differences between having the presenter be physically \newtext{and interactively} ``present'' in AR with the visual content and audience in co-located augmented presentations (i.e., symmetric presenter modality) or be interactively separated from the audience through the use of a tablet (i.e., asymmetric presenter modality).

\subsection{Symmetric vs Asymmetric Mixed Reality Systems}

Mixed reality has been demonstrated for use in multi-user contexts beyond presentation, such as in education~\cite{MONAHAN20081339, phon2014education,Calandra2021training, drey2022collaborative} and remote assistance~\cite{Gurevich2015DesignAI, Gurevich2012TeleAdvisor}.
\newtext{
Multi-user systems often have some form (or forms) of asymmetry in them as different users can have different experiences depending on the system's design~\cite{voida2008asymmetry}.
}
There are multiple kinds of asymmetry involved in our scenario.
First, there is asymmetry in presentation control because of the users' distinct roles: the presenter has exclusive access to the control interfaces and controls which slides are shown and when.
% and they have the exclusive ability to see and interact with the control interfaces. 
% and they have exclusive access to the control interfaces.
Additionally, because an AR HWD blocks part of the wearer's face, their facial expressions become difficult to observe by others, which may generally hamper social interaction in both of our presentation modalities.
Our asymmetric condition also includes an input-output asymmetry: the presenter interacts with a 2D display to manipulate 3D content.

\newtext{
In mixed reality multi-user contexts, setups with ``reality symmetry''
}
% symmetric setups 
are perhaps the standard approach, wherein all users utilize similar interfaces at the same point of the reality-virtuality continuum (RVC)~\cite{milgram1995, Skarbez2021}.
% \newtext{
% % In this paper, we are particularly interested in how asymmetry in presentation device (i.e., both participants using AR HWDs or the presenter using a tablet) affects both users' presentation experiences.
% }
\newtext{
In a survey of co-located AR collaboration research, Radu et al.~\cite{radu2021survey} emphasized how collaborators should be actively aware of each other's attention and emotions and be able to direct the other user's attention.
}
Symmetric AR systems innately allow collaborators to have a shared awareness of what each other sees so that, in our case, the presenter can easily guide their audience throughout the presentation and the shared space~\cite{kurata2004remotecollab, piumsomboon2019collab, radu2021survey, lopez2017awareness}. 
However, present-day HWDs block users' eyes, acting as a barrier to observing their mental states~\cite{gottsacker2022cues} and attention~\cite{matsuda2021reverse}. Perspective challenges also arise, making it difficult to reference areas of interest to one's peers---especially when the focus is on 3D content and visualizations~\cite{smileyMADEAxisModularActuated2021}. To combat this, Chastine et al.~\cite{chastine2007referencing} suggested providing multi-modal referencing techniques and shared viewpoints in collaborative AR environments.

On the other hand, asymmetric setups \newtext{can have unique advantages in multi-user systems in general~\cite{voida2008asymmetry}} and are gaining interest \newtext{in XR contexts in particular} as they provide ``different means to visualize and interact with virtual content'' ~\cite{Grandi2019Asymmetry}, which can improve overall task effectiveness~\cite{Grandi2019Asymmetry} while still facilitating a high degree of immersion to the non-HWD user~\cite{lee2020rolevr}.
Johnson et al.~\cite{johnson2021referencing} also found that when suitable guidance cues are used, experts do not need to be in the immersive environment when referencing areas of interest---though this was mainly in a remote assistance context. For visualization, Reski et al.~\cite{reski2020hybrid} found that having visualizations that share their position, orientation, and field of view of each user was able to support users' ability to make spatial references in an asymmetric setup. However, the incongruent interfaces and perspectives between users result in a diminished common ground~\cite{saffoTheirEyesTheir2023}, which can reduce workspace awareness~\cite{gutwin2002descriptive} and thus, for example, make it more likely to interrupt a VR user disruptively \cite{gottsacker2021diegetic, kudo2021balancing, mcgill2015dose}---or in our case, an audience member who is deeply engaged with the presentation content.

\newtext{
Quite relevant to our work, Drey et al.~\cite{drey2022collaborative} investigated the effects of symmetric and asymmetric teacher interface modalities on communication and group awareness for co-located users in a pair-learning scenario.
In the symmetric condition, both participants used VR HWDs and had avatars.
In the asymmetric condition, the teacher used a tablet with teaching notes and a live spectator view of the student's VR view.
% move most of this to design guidelines/discussion. this part just about study differences.
% They found that their symmetric condition improved communication and workspace awareness, leading them to suggest using symmetric systems whenever possible.
% They also suggested that in asymmetric VR setups, the tablet user should be able to freely explore and interact in the virtual environment to improve their spatial awareness.
% Furthermore, they suggested providing avatars for users and correcting any physical-virtual mismatch between audio or visual cues that may result from virtual locomotion.
% Last, they recommended explicit signaling within the VE when guiding VR users to look at specific objects.
We identify three key differences between our work and that of Drey et al.: (1) Drey et al.'s asymmetric design is suboptimal for social interaction, as the tablet user lacks an avatar in the virtual environment and has limited interactivity; (2) their setup represents a co-located VR environment where neither the other collaborator nor the physical surroundings are visible; and (3) their study focuses on paired learning, which differs from the presentation context. In contrast, our research explores symmetric and asymmetric AR setups for augmented presentations, allowing both AR and tablet users to see each other and interact with the presentation content. In addition to group awareness, our study is particularly interested in the effects of different system configurations on users' social interactions in co-located AR scenarios.}

% For this reason, we chose to investigate user interactions similar to Drey et al.~\cite{drey2022collaborative} but in the context of AR, where face-to-face interactions could be preserved in both the symmetric and asymmetric setups.In our case, the asymmetric modality requires the presenter to maintain engagement and awareness across disjoint perspectives (i.e., they must engage the audience with real-world social cues and with AR presentation content that the presenter cannot view directly in the real world).

Our work, therefore, considers all of these factors across both symmetric and asymmetric mixed reality systems and seeks to understand their influence on audience engagement, social interaction, and shared awareness in co-located augmented presentations.

\subsection{Social Interaction in Cross-Reality}
While recent events have led to a shift towards hybrid or fully remote meetings, people may still prefer in-person communication---especially to help establish more trust~\cite{WILSON200616, Toth2014Trust} and better outcomes~\cite{bradner2002, singh2014persuasion} in high stakes discussions and negotiations. As such, an XR presentation system would likely seek to ensure that any conversation and social interaction between the presenter and their audience is as natural as possible, such as by leveraging behaviors standard in social settings as input modalities (e.g., speech, body gestures, eye gaze).
\newtext{For example, Dagan et al.~\cite{dagan2022irl} showed that designing co-located, symmetric mobile AR games to include shared experiences leveraging face-to-face social signals can foster social engagement.}
Moreover, presenters may eventually need to conduct regular immersive presentations in their day-to-day operations, which may prove fatiguing should an HWD be worn for long periods of time~\cite{biener2022quantifying}. Interpersonal communication may also be negatively affected, such as by \changetext{occluding the presenter's} facial expressions and eye contact \changetext{from the audience's view}~\cite{viola2022seeing, mcatamney2006examination}. For these reasons, it may be that asymmetric presentations, despite any possible downsides compared to symmetric ones, are ultimately more attractive and practical for the presenter than putting on a HWD. However, it remains an open question whether the presenter and audience occupying different ends of the RVC influence the overall social interaction and interpersonal engagement~\cite{slater2009presence, numan2022collab}.

\newtext{
Recent research 
% comparing symmetric and asymmetric collaborative configurations 
highlights the complexities of fostering social engagement, cooperation, and connectedness when participants experience differing levels of immersion and interaction modalities.
For games on traditional 2D screens, Harris and Hancock~\cite{harris2019asymmetry} have shown that designed interdependence in asymmetric cooperative games can improve players' perceptions of connectedness and social engagement over symmetric games.
% To overcome related communication barriers, researchers have investigated sharing physiological signals between wearable users who are separated by significant distance~\cite{liu2021otter} or between VR users who cannot directly observe each other's typical social behaviors~\cite{dey2017physiological} to enhance the quality of their social connection.
However, when comparing symmetric and asymmetric interfaces in a co-located 3D co-manipulation task using VR HWDs and AR tablets, Grandi et al.~\cite{Grandi2019Asymmetry} found that asymmetric interfaces led to high cooperation but lower perceptions of mutual assistance.
Additionally, asymmetric setups in which one user is immersed in XR can make XR users feel isolated and surrounding users feel excluded from the experience~\cite{gugenheimer2018facedisplay}.
To address these challenges, researchers have provided additional channels for the non-immersed user to see and interact with the XR user's virtual environment to facilitate social interaction~\cite{gugenheimer2018facedisplay, jansen2020share}.
Additionally, restoring social cues occluded by the XR HWD --- such as displaying the XR user's eyes on the outside of the HWD --- can improve social interactions~\cite{bozgeyikli2024googly, chan2017frontface, mai2017transparent, matsuda2021reverse}.
Additionally, Dey et al.~\cite{dey2017physiological} showed that sharing (typically invisible) physiological signals between VR users who cannot directly observe each other's social behaviors can enhance the quality of their social connection.
}

\changetext{Our work examines how symmetric and asymmetric presentation styles influence interpersonal dynamics in co-located AR environments to help inform the design of augmented presentation systems that facilitate engagement and natural social interaction.
}

\begin{figure*}[t!]
    \centering
    \newcommand{\subfigwidth}{0.43\linewidth}
    \begin{subfigure}{\subfigwidth}
        \centering
        \includegraphics[width=\linewidth]{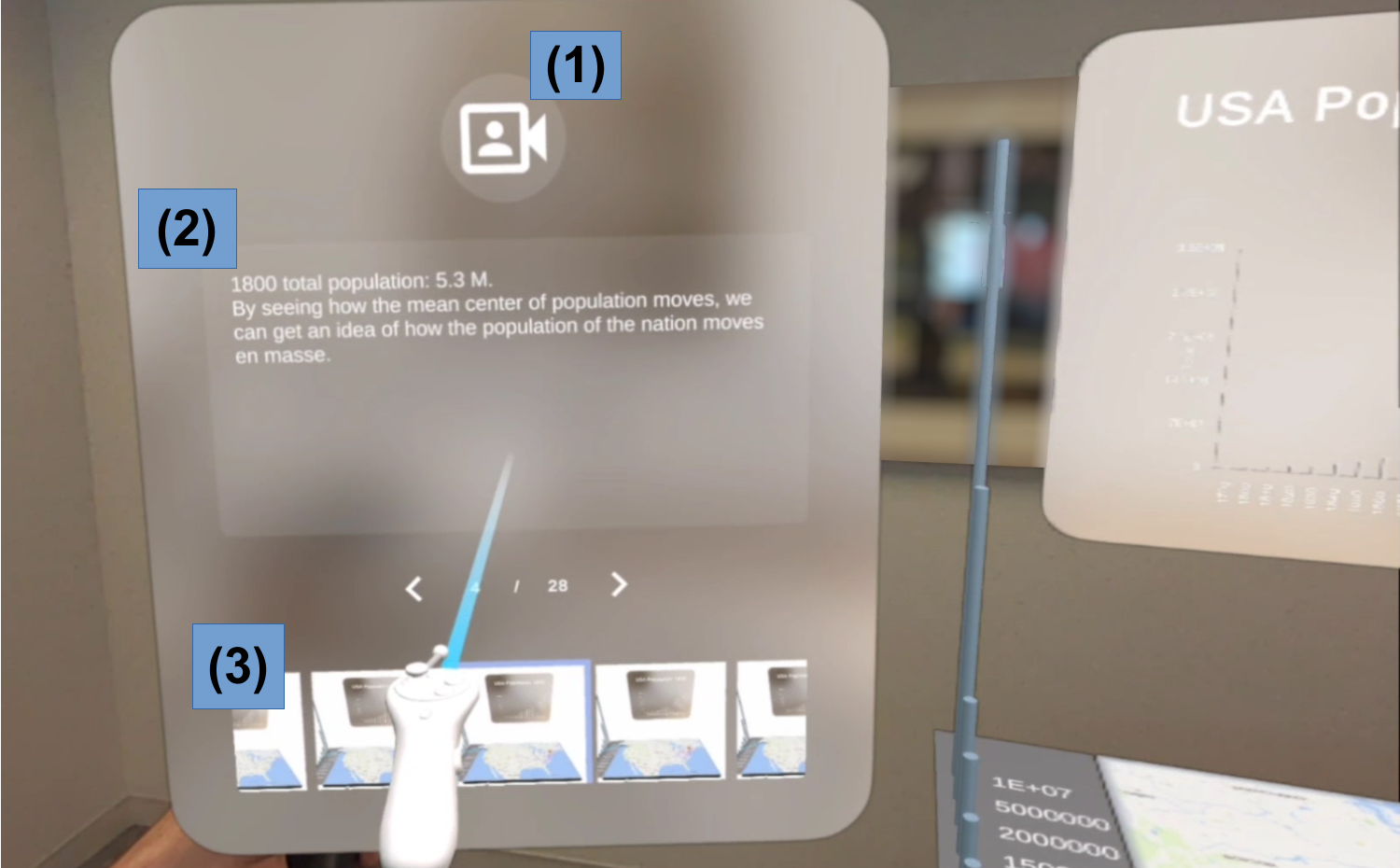}
        \caption{}
        \label{fig:arui}
    \end{subfigure}
    \begin{subfigure}{\subfigwidth}
        \centering
        \includegraphics[width=\linewidth]{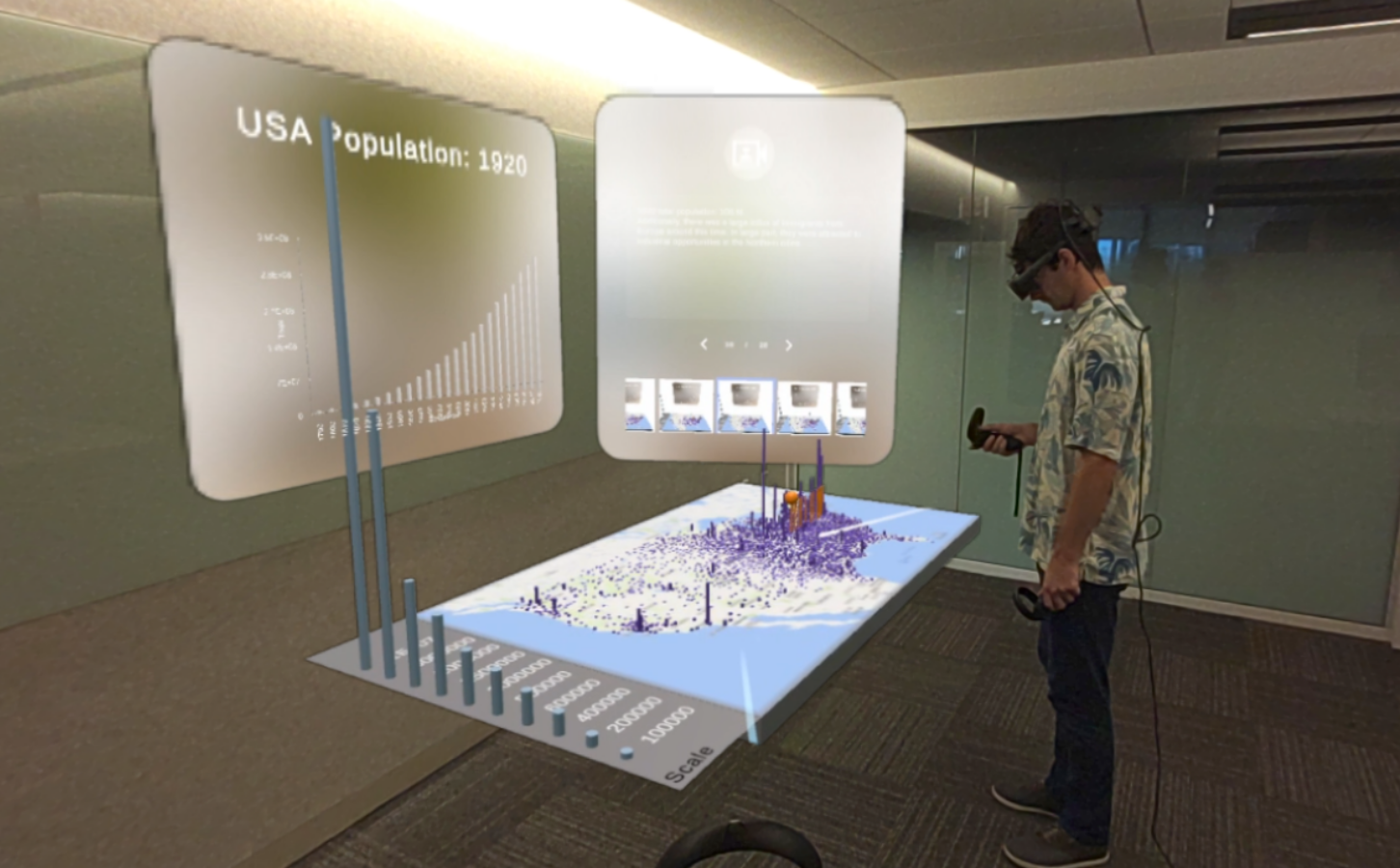}
        \caption{}
        \label{fig:arpoint}
    \end{subfigure}
    \begin{subfigure}{\subfigwidth}
        \centering
        \includegraphics[width=\linewidth]{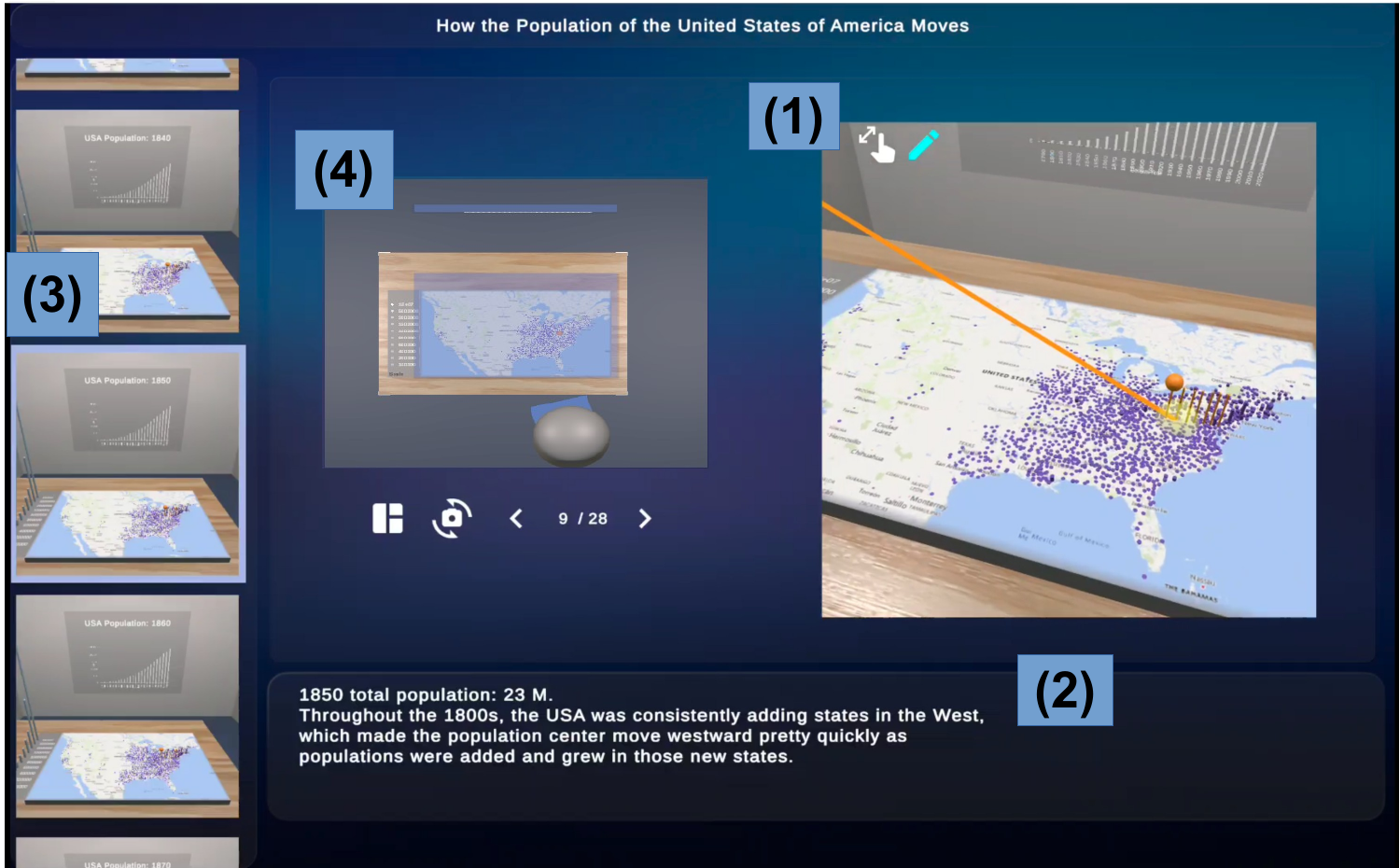}
        \caption{}
        \label{fig:tabletui}
    \end{subfigure}
    \begin{subfigure}{\subfigwidth}
        \centering
        \includegraphics[width=\linewidth]{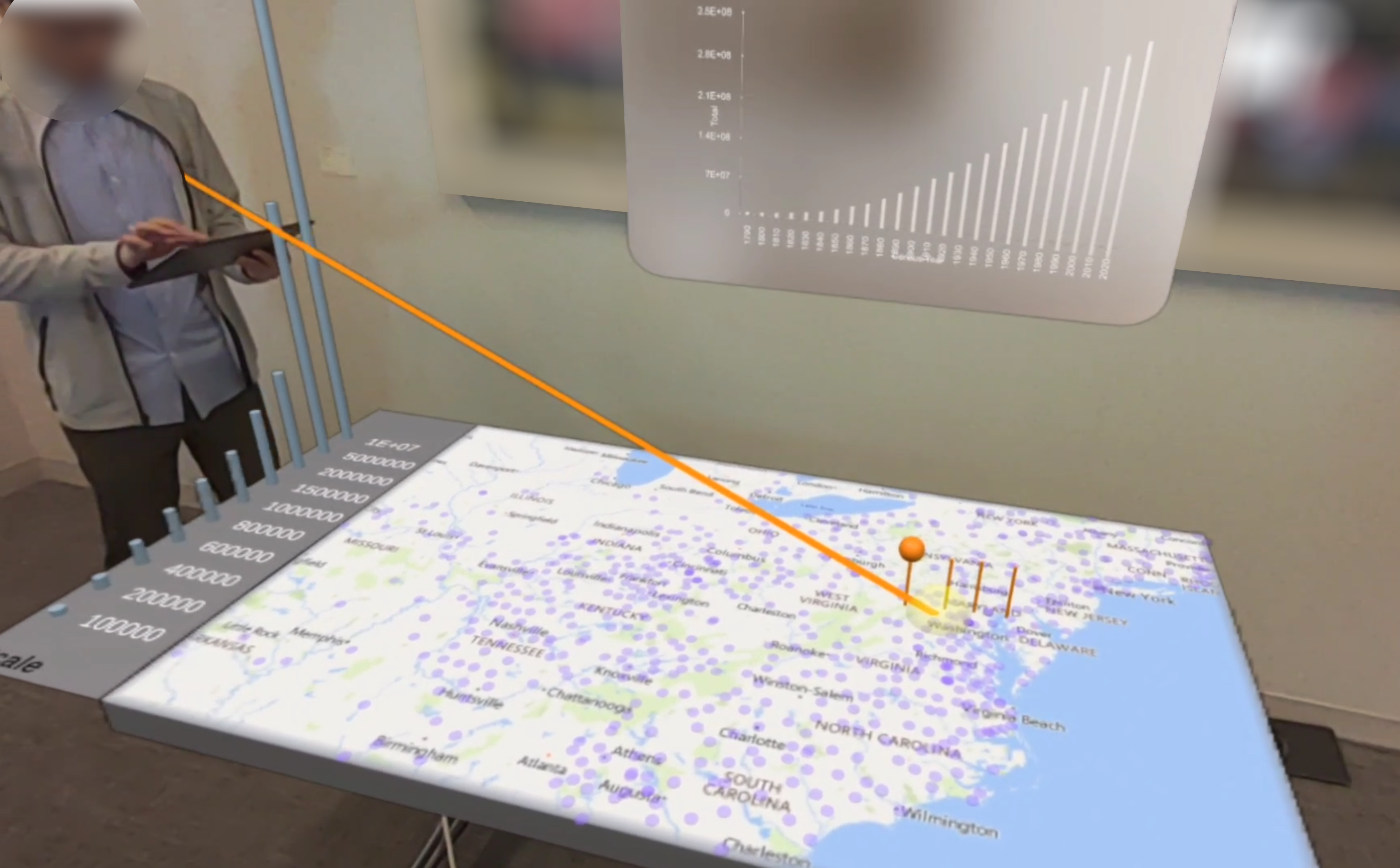}
        \caption{}
        \label{fig:tabletpoint}
    \end{subfigure}
    % Descriptions:
    \Description{(a) A virtual window floats in front of a user's first-person view. From top to bottom, the window contains a video camera icon, a text box with speaker notes, arrow buttons to advance presentation scenes, and a scrolling view of screenshots showing presentation scenes. A virtual controller points at the middle of this window.
    (b) A third-person view of a user wearing an AR HWD standing in front of a virtual map of the United States with thousands of skinny purple cylinders depicting county populations with their height, a virtual window showing a bar chart, and the presenter's control menu. The user is pointing at the map.
    (c) A tablet interface with a vertically-scrolling presentation scene view on the left with screenshots of presentation scenes. In the middle of the interface, there is a top-down view of the presentation space showing an avatar head positioned in the lower right corner and angled slightly to the left. To the right of this is a simulated view of the audience looking at the map. In this view, an orange ray is pointed at the east coast of the USA. Below the top-down and live views on the interface, a text box shows speaker notes.
    (d) A first-person view of an audience member viewing the map with a geospatial data visualization displayed horizontally and a bar chart displayed vertically behind the map. A presenter holding a tablet stands next to the map. An orange ray comes out of the presenter and connects with the east cooast of the USA.
    }
    \caption{Screenshots from our augmented presentation system in use. 
    \textbf{(a)} Symmetric presenter interface (close-up view).
    Features:
    (1) Button to toggle a mirrored view showing a simulated view from the audience's perspective. When activated, the simulated mirrored audience view appears directly above the pictured interface.
    (2) Speaker notes to guide the presenter on the presentation.
    (3) Scene switcher with a carousel view of previews of each scene in the presentation.
    \textbf{(b)} ``Zoomed out'' presenter's view of the presenter interface, the presentation content, and the audience.
    \textbf{(c)} Asymmetric presenter interface for the tablet.
    Features:
    (1) Simulated mirrored audience view.
    The presenter can tap on the virtual content to cast a virtual pointer ray.
    The presenter can press the hand icon button to switch to map interaction mode, allowing them to zoom the map by pinching with two fingers or pan the map by moving one finger.
    (2) Speaker notes.
    (3) Scene switcher.
    (4) Top-down view with simulated audience head to visualize the audience's position and orientation relative to the presentation content.
    The presenter can tap the presentation surfaces shown in this view (the map and the slide behind the map) to stabilize the simulated camera view so that it focuses on these objects.
    \textbf{(d)} Audience's perspective of the presenter using the virtual pointing ray when using the asymmetric presentation modality.
    }
    \label{fig:presentationfigs}
\end{figure*}

\changetext{\section{Prototype Design \& Implementation}}

% \newtext{
% AR is a promising technology for providing engaging experiences for users viewing and interacting with visualizations and other presentation content~\cite{bravo202immersive, saffoIASpace, zhang2024jigsaw}.
% }
% We implemented a custom system for structured 2D and 3D interactive content presentations as a test bed for augmented presentations. 
% Our system enables one or many audience members to experience the augmented presentation in augmented reality using a head-worn display (AR HWD) and hand-held six-degree freedom controllers.
\changetext{
% Our research explores the trade-offs in social interaction and group awareness between different versions of a presenter's control interface.
In this section, we describe the creation of our augmented presentation system prototype, which was used to investigate how symmetric and asymmetric presenter modalities affect the presentation experience of both the presenter and audience.
}

\newtext{\subsection{Augmented Presentation System Design Principles}}

\newtext{
We based the design of our augmented presentation system on related research on co-located AR and 
% high-level 
design goals Kumaravel et al.~\cite{kumaravel2020trasceivr} derived from formative interviews about asymmetrical interactions between VR users and users of non-immersive displays.
First, we aimed to support \textbf{independent exploration} for both users~\cite{kumaravel2020trasceivr}.
While it is a structured presentation where the presenter controls the main flow of the presentation, the audience can feel more engaged through interactivity~\cite{shatri2022interactive, haryani2021impact, ruado2024enhancing, grzeskowiak2015enhancing, stoevesandt2020interactive}.
To this end, we enabled both users to interact with the presentation content.
% This interaction took slightly different forms for each presentation modality.
}

\newtext{
We also aimed to enable both users to make \textbf{efficient and direct spatial references} to presentation content~\cite{kumaravel2020trasceivr, chastine2007referencing, radu2021survey}.
This is required to enable the presenter and audience to discuss specific aspects of the presentation content.
The symmetric setup naturally supports this goal as both users can use their hands to point to the shared AR content.
The asymmetric condition, however, requires additional features to enable the presenter to make and understand references made in the AR space~\cite{saffoTheirEyesTheir2023, johnson2021referencing} (e.g., mirrored views and attention guidance tools).
Further, we aimed to provide \textbf{stable presentation content} for both users~\cite{kumaravel2020trasceivr}.
Simply, stability helps users make precise references to content when pointing to it.
Again, this is straightforward for the symmetric modality because both presenter and audience view the AR content registered to the real world.
% However, since the asymmetric presenter modality requires real-time awareness of where the audience is looking, the presenter needs an option to stabilize their view of the scene so they can make precise references.
To make precise references when using the tablet, the presenter needs a stabilized method of viewing the presentation content instead of through an unsupported hand-held view that is difficult to stabilize or through a view controlled only by the audience.
}

\newtext{
Last, we aimed to support  \textbf{social interaction and co-presence} between users by providing social cues such as body language~\cite{kumaravel2020trasceivr, radu2021survey}.
Expressing and observing social cues is an important aspect of establishing and gauging interest during presentations~\cite{murali2021affective}.
The asymmetric and symmetric modalities have different strengths when it comes to meeting this principle, and investigating the trade-offs between them is indeed the objective of study in this paper.
In particular, the symmetric modality supports a sense of shared context because both users interact in the same workspace.
However, the AR HWD can obscure half of the presenter's face, blocking social expressions and potentially reducing interpersonal connection~\cite{mcatamney2006examination}.
Therefore, the tablet may be better suited for supporting social interaction between users.
}

\subsection{Presentation Interfaces}
We designed and implemented separate interfaces for the audience and symmetric and asymmetric presenter modalities to best suit each device's visual and interaction affordances. 

\paragraph{\textbf{Audience Interface:}}
The audience interface showed the presentation content situated in the real world through the AR HWD.
% Our presentation scenario, in particular, involved a geospatial data visualization on a 3D map registered to a real table and a bar chart on a flat panel against a physical wall.
The audience's controllers included standard ray interactors common in 3D environments~\cite{laviola20173d}.
\newtext{Similar to other research involving user interactions with maps in immersive environments~\cite{yang2018maps},} the audience could pan the map by pointing at it with one of their controller rays, pressing the grip button, and moving the controller in the direction they wanted to pan the map.
\newtext{As an analog to zoom interaction on common map applications on mobile devices,} they could zoom the map by pointing both controllers at the map, pressing both controllers' grip buttons, and moving the controllers toward each other to zoom out or away from each other to zoom in.

\paragraph{\textbf{Symmetric Immersive Interface:}}
The symmetric presenter interface is shown in~\autoref{fig:arui}.
It included the following features \newtext{inspired by presenter interfaces for common desktop presentation software}: a scene-switching scrolling view to allow the presenter to advance the presentation and preview scenes not currently shown, speaker notes to guide the presenter's speech about the current presentation scene, and a button to toggle a simulated mirrored view that shows the audience's current perspective of the presentation content.
%\newtext{The presenter could use rays emanating from their controllers to interact with these virtual controls, which is a common and intuitive interaction method for selection in 3D~\cite{laviola20173d}.}
The presenter could point with the ray interactors on their controllers and interact with the map in the same ways as the audience.
% \note{TODO: justify why the rays were shown when presenter interacted with the control interface.}
Based on findings from Piumsomboon et al.~\cite{piumsomboon2019awareness} showing that persistent head gaze cues improved awareness in AR collaboration, the rays emanating from the audience's and presenter's controllers were always visible to each other to provide continuous interaction awareness cues.
% , similar to how continuous head gaze cues improved awareness in an AR collaboration study by Piumsomboon et al.~\cite{piumsomboon2019awareness}.
% even when the presenter interacted with the presentation control interface (which was invisible to the audience).
% While this may have appeared distracting at times, this was to ensure that users could always see what the other person was gesturing toward or interacting with.

\paragraph{\textbf{Asymmetric Non-Immersive Interface:}}
The asymmetric presenter interface is shown in~\autoref{fig:tabletui}.
It included the same features as the symmetric interface, with additional features supporting group awareness between the presenter and the audience \newtext{inspired by related research~\cite{radu2021survey, saffoTheirEyesTheir2023}}.
\newtext{To support a sense of shared space and provide awareness of the audience's location and attention, the tablet} provided a simulated top-down view of the room, showing the presentation content and the audience's head position and orientation.
\newtext{
The audience's simulated head in this view consisted of a simple capsule object with an attached cube to indicate the direction the audience is facing, similar to an avatar with a view frustum visualization (e.g., ~\cite{sasikumar2019wearable, piumsomboon2019awareness}).
}
\newtext{
Research on remote collaboration involving users who do not share the same task space has shown that sharing users' views can provide common ground and improve referencing communication~\cite{johnson2021referencing, chastine2007referencing, fussel2000coordination, fussel2003remote, kraut2003visual, saffoTheirEyesTheir2023}, and some systems have provided multiple views of the task space to provide more common ground~\cite{rio2018telementoring, rasmussen2019scenecam, johnson2021referencing}.
Therefore, the tablet interface} also provided a mirrored audience view panel, similar to the feature in the symmetric interface, which shows a simulated feed of the audience's view of the presentation content.
\newtext{Similar to effective attention guidance methods in remote-expert collaboration systems~\cite{lee2020viewsharing},} the presenter could point to presentation content by tapping in the mirrored view panel.
In the audience's 3D view, a virtual ray would connect with this pointing target (\autoref{fig:tabletpoint} shows what this looked like from the audience's perspective).
\newtext{An alternative we had considered to support referencing for the presenter was to provide a tablet AR view of the scene that they could use to directly point to things with their real hands or with a virtual ray.
In our initial design phase, we decided against this approach as we found it unwieldy for presenters to hold up the tablet to perform this interaction simultaneously with other presentation controls.}
Additionally, the presenter was able to select a presentation surface in the top-down view, and the mirrored view panel showed a camera that was stabilized and focused on that surface.
This was designed to allow the presenter to point to specific parts of the presentation content or manipulate the map more precisely (i.e., without being affected by the audience's movements in their live feed).

\newtext{To maximize interaction learnability and familiarity, the presenter could interact with the map with common map interactions on touchscreen devices:}
they could pan the map by dragging one finger on the map, and they could zoom the map in or out by using two fingers and separating or pinching their fingers, respectively.
% The presenter could pan the map by dragging one finger on the map.
% They could zoom the map in or out by using two fingers and separating or pinching their fingers, respectively.

\changetext{
\subsubsection{Hardware and Setup}
We built the presentation system using Unity version 2022.3.4.
The system consisted of four applications: an audience application that ran on an AR HWD, a symmetric presenter application that ran on an AR HWD, an asymmetric presenter application that ran on a tablet, and a server application.
The audience application and the symmetric presenter application were run on Dell Precision 7680 laptops each equipped with an Intel i9 CPU, an NVIDIA RTX 4000 Ada Generation Laptop GPU, and 64 GB of RAM.
Each laptop was connected to an HTC VIVE XR Elite HWD with a VIVE Streaming Cable.
% The HTC VIVE XR Elite has a resolution of $1920\times1920$ per eye, a refresh rate of 90 frames per second, a diagonal field of view of 110$^{\circ}$, and accommodates interpupillary distance (IPD) in the range of 54 to 73 mm.
The asymmetric presenter application was run on a Samsung Galaxy Tab S8+.
The server application relayed networking messages between the presenter and audience to synchronize presentation scene data and user position data.
Networking was implemented with Mirror Networking\footnote{\url{https://github.com/MirrorNetworking/Mirror}}.
}
% Manual coordinate-space calibration procedure.

\section{\changetext{User Study}}
% \todo{Consider restructuring...many sub-sub-sections.}
We believe augmented presentation has the potential to benefit the experience and outcomes of co-located one-on-one presentations. 
To further examine this premise, we constructed and conducted a user study using our system for augmented presentations. 
This work examines the effects of augmented presentation and symmetric or asymmetric presenter modalities on engagement, group awareness, and social interaction. 
This section details the study design, implementation, and execution of our within- and between-subject user study, designed to satisfy this goal. 

% \subsection{Methods}
We designed a within- and between-subject user study in which participant pairs utilized our augmented presentation system to experience or present a pre-authored presentation. 
We collected qualitative behavioral data, self-reported questionnaires, and semi-structured participant interviews. 
The presenter control interface was designed within subjects where each presenter presented twice, once using the symmetric immersive interface for one and the asymmetric non-immersive interface for the other. 
The order in which the interface was used first was counterbalanced among presenters. 
This design was chosen to allow us to glean insights from presenters on the differences and preferences between control interfaces.
The audience experience was designed between subjects. Each audience participant saw the presentation once with a presenter using one of the two control interfaces. 
This design was chosen to allow us to observe differences in audience behavior and responses with each control interface.

\changetext{
\subsection{Presentation Content}
\label{sec:pres_content}
Using our system, we designed and implemented a narrative data-driven presentation incorporating static and interactive elements \newtext{in 2D and 3D, as both are prevalent formats of information presentation and consumption in today's presentation/exhibition experiences~\cite{stjohn2001shape, tory2006viz, dubel20142d3d, gonccalves2016not}.
% depending on the underlying data and what users are trying to understand about it.
}
We chose US census data as the presentation topic, as it is well suited for spatial presentation, visualization, interaction, and is straightforward and relevant for many people even without prior background or expertise.
}
% When designing our presentation content, we aimed to find a topic well suited for spatial presentation, interactivity, and engagement with a wide range of audiences. 
% Furthermore, we wanted the content to be straightforward for people seeing it for the first time, \newtext{with no prior background or expertise}. 
% With this in mind, we focused the presentation content on US census data, particularly population migration over time, for the following reasons. 
% Geographic data visualization is well-suited for 3D presentations due to its natural spatial mapping and its metaphors of physical maps. 
% Furthermore, migration data allows us to build a visualization narrative using space and time on a topic familiar to our target audience of US-based participants. 
% Finally, this data and corresponding historical context can be summarized at a high level, making it easy to learn and present quickly.
\changetext{
Using this data, we created a presentation about the history of population growth and migration in the US from 1790 to 2020, inspired by a narrative visualization video by Danielle Kurtzleben and Vox\footnote{\url{https://www.youtube.com/watch?v=OTaYKEumwc8}}. 
The presentation was divided into slides for each decade of our time range.
\newtext{Each slide in the presentation contained a map visualizing this data in 3D}, and a \newtext{2D bar chart} visualization of the total population by decade.
\changetext{Each slide also had speaker notes as a reminder to the presenter of the changes in the data across each decade.}
The map was spatially registered to a real table. The population data was visualized on this map with 3D cylinders for each county, population values being encoded by height.
Furthermore, orange cylinders encoded the population mean center\footnote{\url{https://www.census.gov/geographies/reference-files/time-series/geo/centers-population.html}} for each decade to further emphasize the shift in the US population over time. 
% The presenter and audience members can control the interactive map to pan and zoom the map with the data from the current slide.
}

\subsection{Measures}

We used a combination of self-report questionnaires, video recordings of experiment sessions, and semi-structured interviews to examine participants' presentation experiences in terms of engagement, group awareness, social interaction, and overall experiences.

\paragraph{Questionnaires}
\label{sec:methods:q}
\begin{itemize}
    \item \textbf{Cognitive Load.} We used the Mental Demand, Physical Demand, and Performance sub-scales of the NASA Task Load Index (NASA-TLX)~\cite{hart2006tlx} to measure aspects of presenters' self-assessed cognitive load during the presentation experience.
    Mental Demand refers to ``How much mental and perceptual activity was required (e.g., thinking, deciding, calculating, remembering, looking, searching, etc.)''~\cite{hart2006tlx}.
    Physical Demand refers to ``How much physical activity was required (e.g., pushing, pulling, turning, controlling activating, etc.)''~\cite{hart2006tlx}.
    Performance refers to how successful participants thought they were in accomplishing the goals of the experimental task~\cite{hart2006tlx}.
    We used a 7-point Likert scale for the questions (1 = low workload/poor performance, 7 = high workload/good performance).
    % The performance sub-scale was inverted for analysis.

    \item \textbf{Group Awareness.} Similar to Schroeder et al.~\cite{schroeder2023dyadic} we adapted the descriptive questions from Gutwin and Greenberg's framework~\cite{gutwin2002descriptive} to formulate questions about users' perceptions of group awareness.
    We created questions for the \textit{what} and \textit{where} categories, but we left out the questions about \textit{who} because the participants were only interacting with one other person.
    We included two questions about artifact awareness from the perspectives of both the participant and their partner.
    The questions asked participants to rate on a 7-point scale how much they agreed (1 = strongly disagree, 7 = strongly agree).
    The specific statements are shown in the supplementary materials.
    % ~\autoref{tab:awareness}.

    \item \textbf{Presentation Engagement.} We used a questionnaire created by Webster and Ho~\cite{webster1997engagement} measuring overall presentation engagement.
    We used a 7-point Likert scale for the questions (1 = strongly disagree, 7 = strongly agree).
    The questions were adapted for both audience and presenter, and they are shown in the supplementary materials.
    % ~\autoref{tab:engagement}.
\end{itemize}

\paragraph{Video recordings.}
We recorded videos of the participants in each presentation session and annotated occurrences of the following behaviors:
\begin{itemize}
    \item \textbf{Social interactions.} We annotated each instance of participants gesturing to each other, nodding at each other, smiling at each other, and laughing with each other. We also annotated the beginning and ending timestamps for every instance in which participants explicitly looked at each other.
    % \begin{itemize}
    %     \item Social expressions: We annotated each instance of participants gesturing to each other, nodding at each other, smiling at each other, and laughing with each other.
    %     \item Time looking at partner: We annotated the beginning and ending timestamps for every instance in which participants clearly looked at each other.
    % \end{itemize}

    \item \textbf{Referencing behaviors.} We annotated each instance of participants establishing references with each other.
    Based on related studies examining how groups share awareness and make references in collaborative asymmetric settings~\cite{chastine2007referencing}, we differentiated between the following referencing behaviors: when a participant \textit{pointed} to presentation content, when a participant used \textit{deictic speech} (i.e., spatial language often accompanying pointing such as, ``this'' or ``that''), when a participant used \textit{relative language} (i.e., describing a virtual object in relation to another, e.g., ``to the left of the map''), when a participant used a \textit{property reference} (i.e., describing a virtual object based an attribute, e.g., ``the orange cylinder''), and when a participant used referential chaining (i.e., describing a virtual object based on a previous reference, e.g., ``to the left of the last one'').
    % To characterize how participants established references with each other, we annotated instances of the following behaviors:
    % instances in which participants pointed to specific presentation content to communicate 
    % \begin{itemize}
    %     \item Pointing to virtual content: when a participant pointed to certain content to communicate something to their partner.
    %     \item Deictic speech: spatial language often accompanying pointing such as, ``this'' or ``that.''
    %     \item Relative speech: describing a virtual object in relation to another, e.g., ``to the left of the map.''
    %     \item Property reference: describing a virtual object based an attribute, e.g., ``the orange cylinder.''
    %     \item Referential chaining: describing a virtual object based on a previous reference, e.g., ``to the left of the last one.''
    % \end{itemize}
\end{itemize}

\paragraph{Interview transcripts.}
In the semi-structured interview, we asked participants questions in the following categories to provide further insight 
% into their questionnaire scores and behaviors 
about the effects of the different presentation modalities:

\begin{itemize}
    \item \textbf{Group Awareness.}
    We asked participants to describe how they referenced virtual objects and directed their partner's attention during the presentation.
    We asked them to describe what was straightforward or complicated about making/understanding references and if they could think of any alternative methods they would want to help with this process.
    We asked presenters whether they experienced differences in their abilities to make/understand references with each presenter modality.
    % \begin{itemize}
    %     \item Please describe how you made references to virtual objects, or how you directed your partner’s attention to specific virtual objects when (you/the presenter) was using the tablet to control the presentation.
    %     \item What was straightforward/difficult for you when referring to objects?
    %     \item Can you think of an alternative method you want to help you make references?
    % \end{itemize}

    \item \textbf{Engagement.}
    We asked participants to describe what was most engaging about the presentation and when they felt most engaged.
    We also asked audience members whether they felt more engaged with the presentation content or the presenter themselves and how they engaged them.
    We asked the presenters what techniques they used for engaging the audience with the presentation content and interpersonally and whether they experienced any differences in their abilities to engage the audience with each presenter modality.
    % \begin{itemize}
    %     \item When were you most engaged with the presentation?
    %     \item Did you feel more engaged with the presentation content or with the presenter? 
    %     \item What was engaging about the presentation?
    %     \item How did the presenter themselves engage you in the presentation?
    % \end{itemize}

    \item \textbf{General Feedback.}
    We asked participants if they had ideas for additional features they would like to see in this augmented presentation system.
    We also asked them if they would use this presentation system and to compare it to traditional presentation systems like PowerPoint.
    % \begin{itemize}
    %     \item What changes or additional features would you like to see made to the presentation systems? 
    %     \item If you had this kind of system, would you use it? 
    %     \item Compare this presentation system to other presentation methods, e.g., PowerPoint. 
    % \end{itemize}
\end{itemize}

% We asked the presenter the complement of the engagement questions, e.g., ``When did the audience seem to be most engaged with the presentation?''
% We also asked the presenter to compare each of these aspects when using each version of the presenter interface.

\subsection{Analysis}

% \paragraph{Self-reported Questionnaires.}
% Due to the nature of this study as exploratory research intended to inform future research on immersive presentation systems, we report descriptive statistics about our quantitative data (i.e., questionnaire responses and behavior observations) without statistical test results~\cite{vornhagen2020stat}.
% We present means and standard deviations of participants' questionnaire responses.

To examine how participants engaged with the presentation, interacted socially, and established group awareness with each presentation system, we aggregated and visualized participant questionnaire responses, annotated behavioral observations in the video recordings of the experimental presentation sessions, and conducted a thematic analysis on the semi-structured interview transcripts.

\paragraph{Questionnaire Responses}
We grouped the presenter responses to the cognitive workload questions by presentation modality, computed the medians across sessions, and visualized the results in box plots.
For the awareness and engagement questionnaires, we first averaged the scores for all questions per participant to arrive at average awareness and engagement scores per participant per session.
Then, we grouped participant responses by role (presenter or audience) and presentation modality before finding their medians and visualizing them in box plots.

\paragraph{Behavioral Observations}
% We present counts of specific participant behaviors during the presentation sessions for the behavior observations.
% To shed light on how participants interacted socially and how participants established group awareness, we coded the video recordings of the experimental presentation sessions.
To code participants' behaviors during the study sessions, researcher R1 drafted a codebook of behaviors to observe based on the study's aims of social interaction and group awareness.
Then, researchers R1, R2, R3, R4, and R5 met to discuss the code book and suggest changes.
The final codebook included codes such as \textit{looking at partner} and \textit{smiling at partner} for social interactions and \textit{pointing to virtual content} and \textit{using deictic speech} for group awareness.
Researchers R1 and R2 each coded three presentation videos, while R3, R4, and R5 each coded two sessions using BORIS software~\cite{BORIS}.
The researchers used the BORIS software~\cite{BORIS} to watch synchronized video recordings of the presenter's view, the audience's view, and a third-person view of both participants in the experimental space.
Using BORIS, they inserted annotations of the codes at the timestamps at which they were observed.
R1 reviewed and validated all coded videos and met with the original coder to resolve disagreements.
We grouped participant' social interaction behaviors by role and presentation modality, computed their medians, and visualized them in box plots.
We followed the same procedure for the total number of references participants made.
We also grouped the data by presentation modality and participant role, and we computed the median number of occurrences of specific reference behaviors per presentation session.

\paragraph{Semi-Structured Interviews}
We performed an inductive thematic analysis~\cite{braun2012thematic} on interview transcripts to draw common themes and conclusions from our participants' responses to our interview questions.
Researchers R1 and R2 independently analyzed the transcripts of the semi-structured interviews to generate codes, then collaboratively reviewed the results and resolved disagreements with all researchers to create the code book.
Next, R1 and R2 coded all of the transcripts \newtext{using the finalized codes}.
Then, R1 and R2 met to resolve any disagreements.
Finally, they formed code categories, organized them into sub-themes, and extracted themes.

\subsection{Procedure}

When a pair of participants arrived, they first signed a consent form.
An experimenter briefed each participant separately on the experimental task and introduced the interface they would use in their assigned experiment condition (e.g., AR HWD for the audience and AR HWD or tablet for the presenter).
The presenter was briefed on the presentation content \newtext{(see~\autoref{sec:pres_content})}.
\newtext{The presenter was given as much time as they needed to become familiar with the presentation interface and content (typically 5-10 minutes) and informed the experimenter when they were ready to present}.
The presenter then gave the presentation to the audience.
After the presentation, the audience participants filled out the experiment questionnaires, and an experimenter conducted a semi-structured interview.
Presenters performed this procedure again with whichever interface they did not use the first time and with a different audience participant.
The order in which presenters used the interfaces was counterbalanced between participants.
Presenters completed the questionnaires and semi-structured interviews after both presentations were complete.

\subsection{Participants}
% \newtext{A power analysis was conducted using G*Power to determine the required sample size for this study. Based on an expected moderate effect size of Cohen's d = 0.5, an alpha level of 0.05, and a desired power of 0.95, the analysis indicated that a minimum sample size of 16 presenter participants and 45 audience participants would be required to detect a statistically significant effect for within-subjects and between-subjects tests, respectively.
% However, due to the exploratory nature of this research, we decided to recruit a smaller number of participants and forego statistical power in favor of a deeper qualitative analysis.}
We recruited 18 participants from our organization, 12 as audience and 6 as presenters, for 12 participant pairs. 
% \newtext{Participants held various roles in our organization, including legal, finance, technology, and client services.}
\newtext{
Our audience participants worked in the following roles in our organization: legal (2 participants), finance (2), data science (1), technology/engineering (7).
Presenters worked in the following roles: finance (1), technology/engineering (2), human resources (3).
}
We asked participants to report their experience with using AR or VR technology, playing video games, and giving presentations to an in-person audience on a 5-point scale (1 = Never, 2 = Occasionally, 3 = Once a week, 4 = Several times a week, 5 = One or more hours per day).
9 participants had experience with mobile AR.
5 participants had experience with head-worn AR.
7 participants had experience with head-worn VR.
For AR and VR experience, participants reported an average of $1.56$ ($SD = 0.96$).
For video game experience, participants reported an average of $2.17$ ($SD = 0.96$).
For presentation experience, participants reported an average of $2.27$ ($SD = 0.93$).
\newtext{More specifically, 4 presenters reported occasionally giving presentations, and 2 presenters reported giving presentations several times a week.}
\newtext{None of our participants knew each other before participating in the experiment.}
\section{\changetext{Questionnaire and Observational Data}}
In this section, we first present the responses to the post-study questionnaire and a summary of participant behaviors we observed.
Due to the nature of this study as exploratory research intended to inform future research on immersive presentation systems,
% and our small sample size ($n = 6$ groups for each experimental condition), 
we report descriptive statistics about our quantitative data without statistical test results~\cite{vornhagen2020stat}.
% \newtext{Unless otherwise noted in this section, we did not find any trends related to presenters' presentation experience.}
\newtext{We report only the most relevant results in this section.
For a more detailed look at the quantitative results, refer to the supplementary materials.
% ~\autoref{sec:appendix}.
}

\begin{figure*}[t]
    \centering
    % \newcommand{\subfigwidth}{0.3\textwidth}
    % First subfigure
    \begin{subfigure}{0.3\textwidth}
        \centering
        \includegraphics[height=21em]{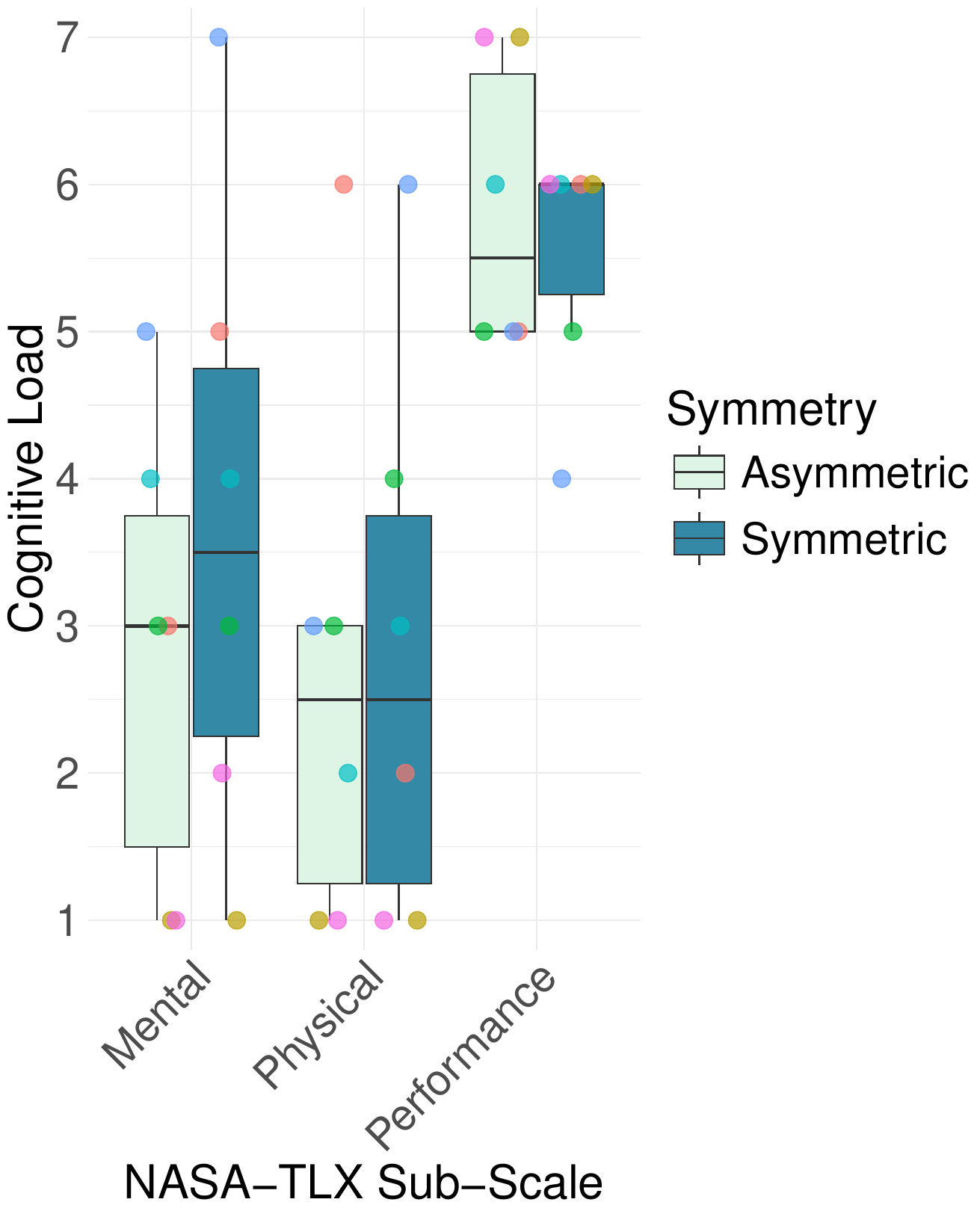}
        \caption{Presenter Cognitive Load}
        \label{fig:tlx}
    \end{subfigure}
    \hspace{6ex}
    \begin{subfigure}{0.18\textwidth}
        \centering
        \includegraphics[height=21em]{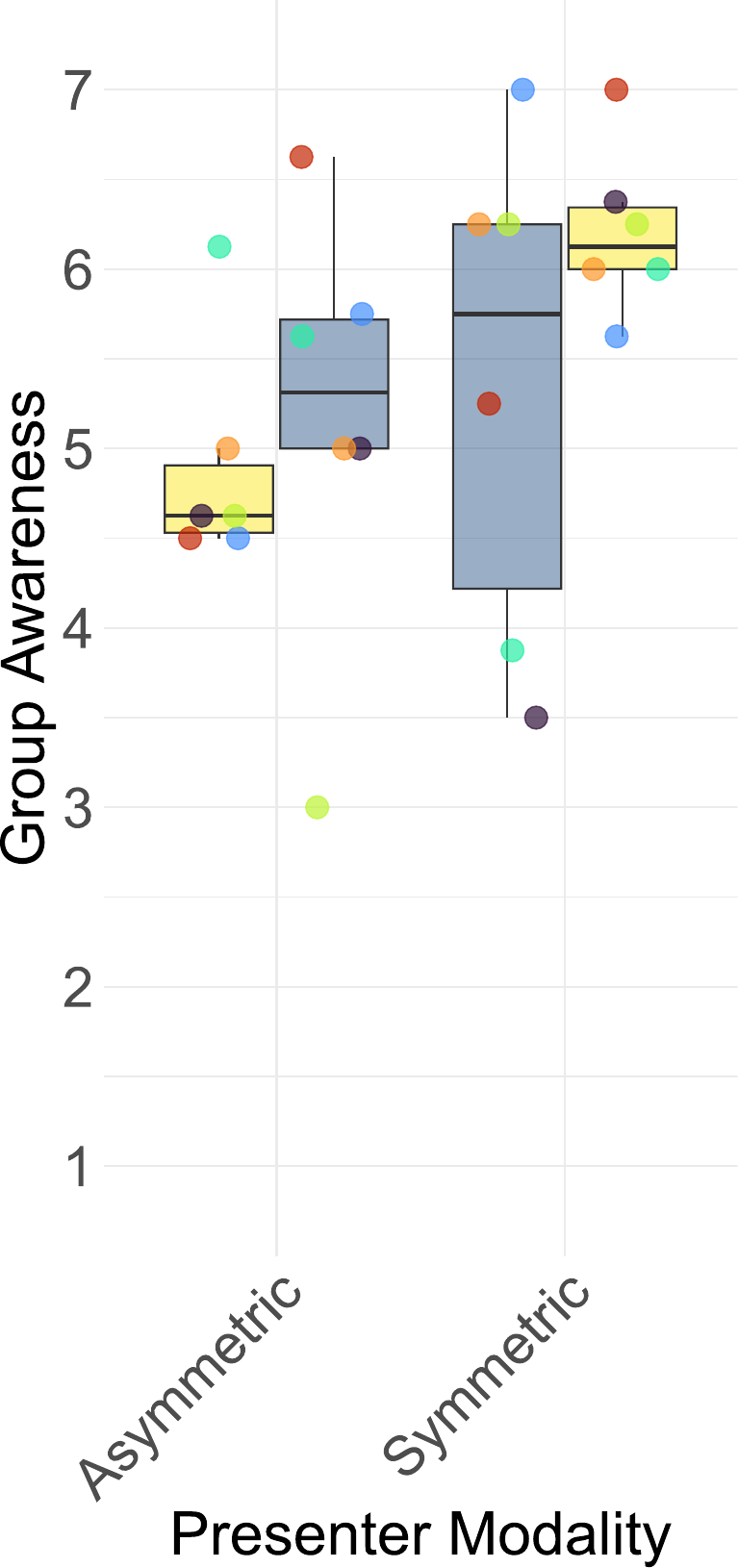 }
        \caption{Group Awareness}
        \label{fig:awareness}
    \end{subfigure}
    \begin{subfigure}{0.28\textwidth}
        \centering
        \includegraphics[height=21em]{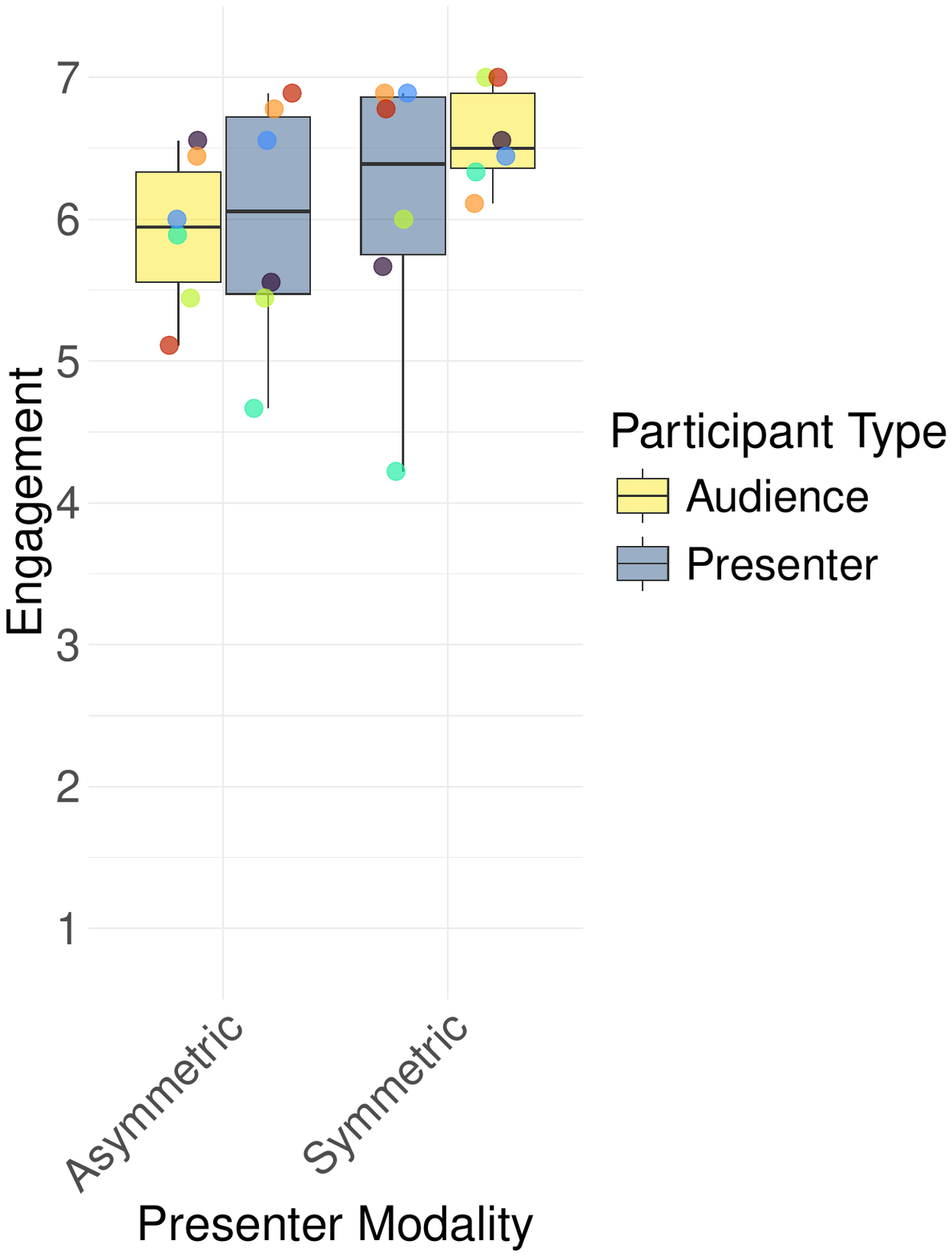}
        \caption{Engagement}
        \label{fig:engagement}
    \end{subfigure}
    % Description
    \Description{(a) Box plot with Cognitive Load on the y-axis and NASA-TLX Sub-Scale on the x-axis. Three groups of x-axis boxes: Mental, Physical, and Performance sub-scales. Two boxes in each pair with different colors: aymmetric as light blue and symmetric as darker blue.
    (b) Box plot with Group Awareness on the y-axis and Presenter Modality on the x-axis. Two groups of x-axis boxes based on Presenter Modality: asymmetric and symmetric. Two boxes in each pair with different colors: Audience as yellow and Presenter as muted blue.
    (c) Box plot with Engagement on the y-axis and Presenter Modality on the x-axis. Two groups of x-axis boxes based on Presenter Modality: asymmetric and symmetric. Two boxes in each pair with different colors: Audience as yellow and Presenter as muted blue.
    }
        \caption{Box plots for questionnaire responses.
        (a) Average presenter scores for the mental demand (lower is easier), physical demand (lower is easier), and performance (higher is better) categories of the NASA-TLX~\cite{hart2006tlx}.
        (b) Average scores for questions about group awareness. 
        % (see~\autoref{tab:awareness}).
        Points represent individual participant responses and are colored by presentation group and presenter: within a presenter modality, points of the same color represent participants in the same presentation session; across presenter modalities, points of the same color represent the same presenter participant.
        (c) Average scores for questions about the presenter's ability to engage the audience in the presentation and the audience's ratings of their engagement in the presentation. 
        % (see the supplementary materials).
        % ~\autoref{tab:engagement}).
        Points represent individual participant responses and are colored by presentation group and presenter.
        }
    \label{fig:questionnaires}
\end{figure*}

\subsection{Questionnaire Responses}
In this section, we report the results of the questionnaires described in~\autoref{sec:methods:q}.
All questionnaire responses are on a 7-point Likert scale.
These results are visualized in~\autoref{fig:questionnaires}.

On the NASA-TLX \textbf{cognitive load} questionnaire, presenters reported a slightly higher mental demand score for the symmetric presentation modality (Mdn.: $3.50$; IQR: $2.50$, $2.25 - 4.75$) compared to the asymmetric modality (Mdn.: $3.00$; IQR: $2.25$, $1.50 - 3.75$).
Presenters rated their \textbf{group awareness} higher in the symmetric presentation modality (Mdn.: $5.75$; IQR: $2.03$, $4.22 - 6.25$) compared to the asymmetric modality (Mdn.: $5.31$; IQR: $0.72$, $5.00 - 5.72$).
Similarly, audience participants rated their group awareness higher in the symmetric presentation modality (Mdn.: $6.12$; IQR: $0.34$, $6.00 - 6.34$) compared to the asymmetric modality (Mdn.: $4.62$; IQR: $0.38$, $4.53 - 4.91$).
Presenters' \textbf{engagement} scores were higher for the symmetric modality (Mdn.: $6.39$; IQR: $1.11$, $5.75 - 6.86$) compared to the asymmetric modality (Mdn.: $6.06$; IQR: $1.25$, $5.47 - 6.72$).
Similarly, audience participants' engagement scores were higher for the symmetric modality (Mdn.: $6.50$; IQR: $0.53$, $6.36 - 6.89$) than the asymmetric modality (Mdn.: $5.94$; IQR: $0.77$, $5.56 - 6.33$).
\newtext{Presenters who reported having weekly experience giving presentations tended to rate their ability to engage the audience more highly with a median score of $6.81$ (IQR: $0.17$, $6.72 - 6.89$) compared to $5.61$ (IQR: $0.94$, $5.25 - 6.19$) for those with occasional presentation experience.}

\subsection{\changetext{Observational Data}}
% section for Boris / observed activities summary
In this section, we report quantitative summaries of the behaviors we annotated in the video recordings of the experimental presentation sessions.
Presenters exhibited more \textbf{social expressions} in the asymmetric modality (Mdn.: $1.50$; IQR: $4.50$, $0.00 - 4.50$) compared to the symmetric modality (Mdn.: $0.50$; IQR: $2.50$, $0.00 - 2.50$). 
Audience participants also exhibited more social expressions in the asymmetric modality (Mdn.: $4.50$; IQR: $13.50$, $0.00 - 13.5$) compared to the symmetric modality (Mdn.: $2.50$; IQR: $1.75$, $1.25 - 3.00$).
We observed that presenters tended to \textbf{look at} the audience for a greater percentage of the presentation when using the asymmetric modality (Mdn.: $6.65\%$; IQR: $8.30\%$, $1.65\% - 9.95\%$) compared to the symmetric modality (Mdn.: $5.65\%$; IQR: $10.05\%$, $1.15\% - 11.20\%$).
However, we observed the opposite tendency for audience participants: that they looked at the presenter more when using the symmetric modality (Mdn.: $5.45\%$; IQR: $8.20\%$, $3.80\% - 12.00\%$) compared to the asymmetric modality (Mdn.: $2.80\%$; IQR: $4.33\%$, $1.35\% - 5.68\%$).
Participants had more occurrences of \textbf{eye contact} in the asymmetric modality (Mdn.: $2.50$; IQR: $3.25$, $0.50 - 3.75$) compared to the symmetric modality (Mdn.: $2.00$; IQR: $3.50$, $1.00 - 4.50$).
Similarly, they spent a slightly longer duration of the presentation making eye contact when using the asymmetric modality (Mdn.: $1.00\%$; IQR: $2.47\%$, $0.18\% - 2.65\%$) compared to the symmetric modality (Mdn.: $0.38\%$; IQR: $2.52\%$, $0.08\% - 2.60\%$).

We observed that presenters tended to make more \textbf{references} in the symmetric modality (Mdn.: $11.50$; IQR: $13.75$, $7.25 - 21.00$) compared to the asymmetric modality (Mdn.: $8.50$; IQR: $6.80$, $4.00 - 10.80$).
Audience participants made references at a similar frequency in both symmetric (Mdn.: $5.50$; IQR: $1.00$, $5.00 - 6.00$) and asymmetric modalities (Mdn.: $7.00$; IQR: $5.25$, $2.50 - 7.75$).
We observed that presenters tended to make more \textit{pointing} references when using the symmetric presentation modality (Mdn.: $6.00$; IQR: $7.45$, $3.75 - 11.20$) compared to the asymmetric modality (Mdn.: $3.50$; IQR: $6.25$, $1.50 - 7.75$).
Audience members made a similar number of pointing references across both symmetric (Mdn.: $3.00$; IQR: $3.50$, $1.25 - 4.75$) and asymmetric modalities (Mdn.: $4.50$; IQR: $4.00$, $1.75 - 5.75$). 
Presenters tended to make more \textit{verbal} references when using the symmetric presentation modality (Mdn.: $7.00$; IQR: $7.75$, $2.00 - 9.75$) compared to the asymmetric modality (Mdn.: $2.50$; IQR: $4.75$, $2.00 - 6.75$).
Audience members made a similar number of pointing references across both symmetric (Mdn.: $3.00$; IQR: $3.50$, $0.50 - 4.0$) and asymmetric modalities (Mdn.: $2.00$; IQR: $2.75$, $0.25 - 3.00$).
Participants mostly used deictic speech for their verbal references.

\begin{figure*}[t]
    \centering
    \newcommand{\subfigwidth}{0.28\textwidth}
    % \newcommand{\subfigwidth2}{0.125\textwidth}
    % First subfigure
    \begin{subfigure}{\subfigwidth}
        \centering
        \includegraphics[height=21em]{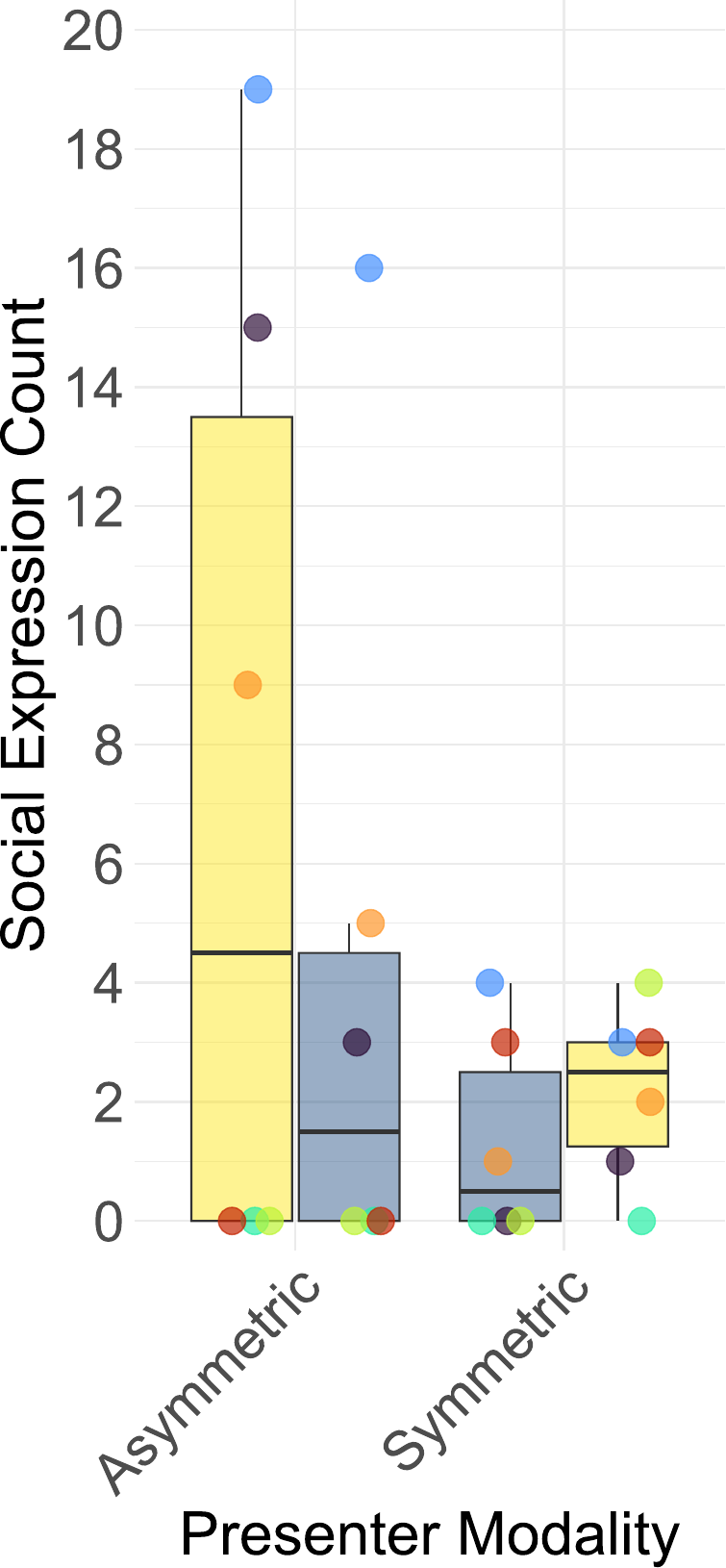}
        \caption{}
        \label{fig:socialexpr_pid}
    \end{subfigure}
    % \hspace{3ex}
    \begin{subfigure}{\subfigwidth}
        \centering
        \includegraphics[height=21em]{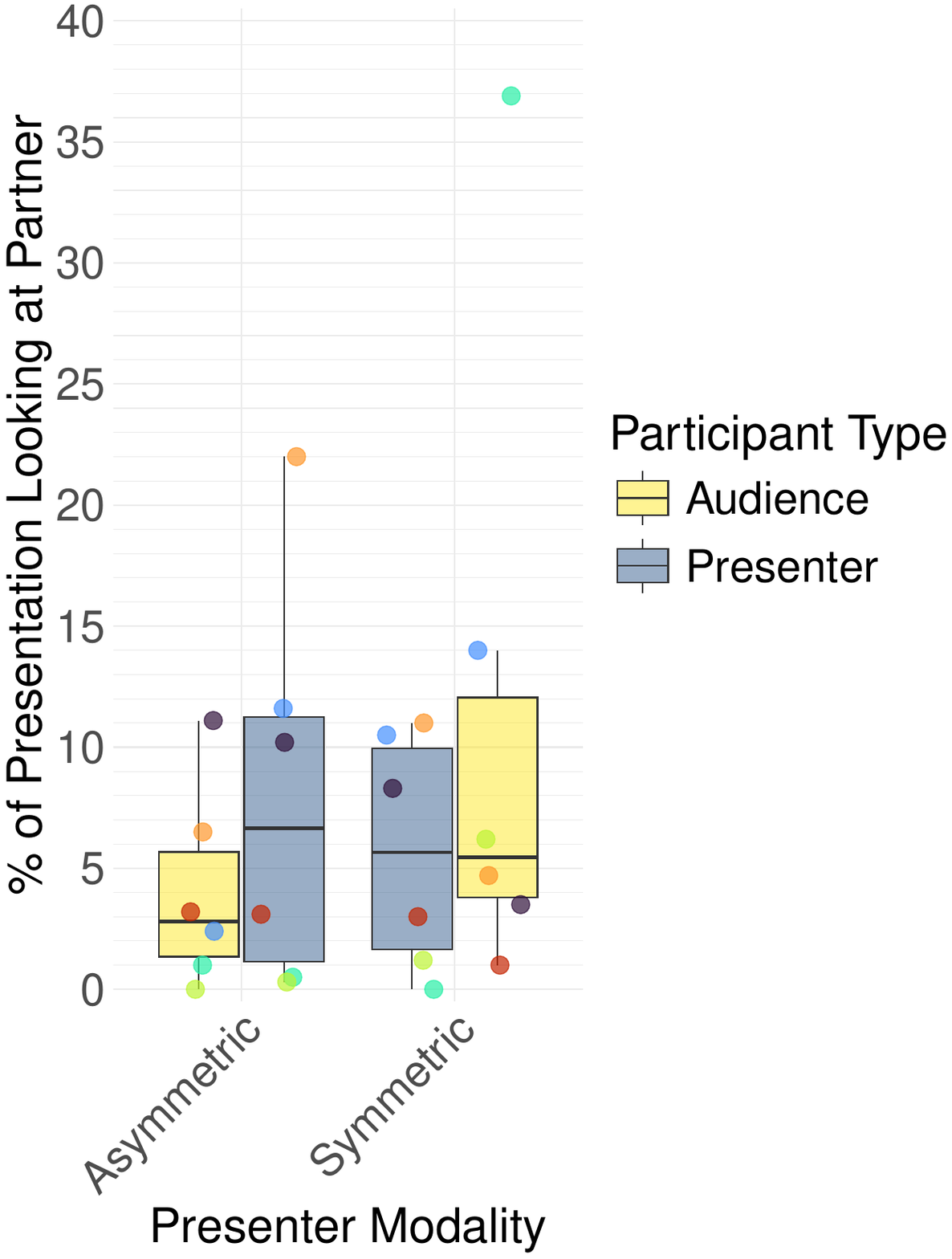}
        \caption{}
        \label{fig:lookat}
    \end{subfigure}
    \hspace{12ex}
    \begin{subfigure}{0.14\textwidth}
        \centering
        \includegraphics[height=20em]{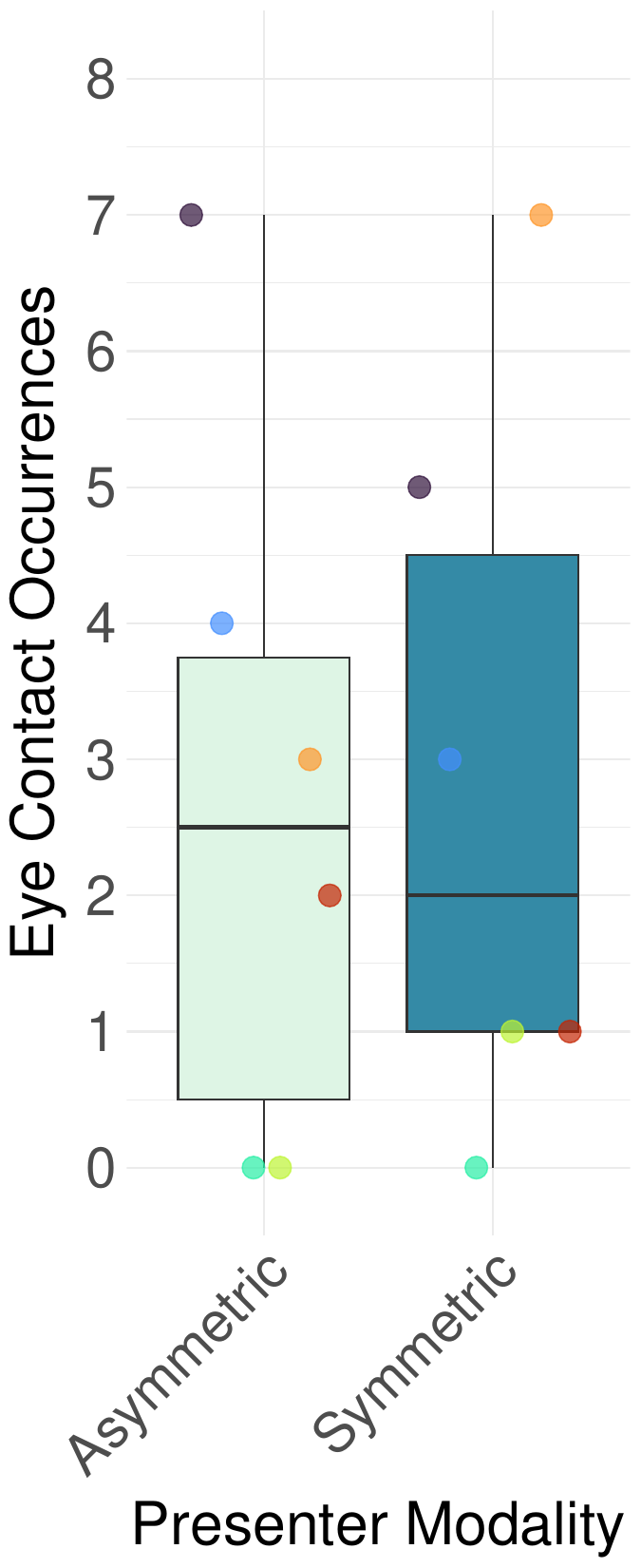}
        \caption{}
        \label{fig:eye_occur}
    \end{subfigure}
    \hspace{3ex}
    \begin{subfigure}{0.14\textwidth}
        \centering
        \includegraphics[height=20em]{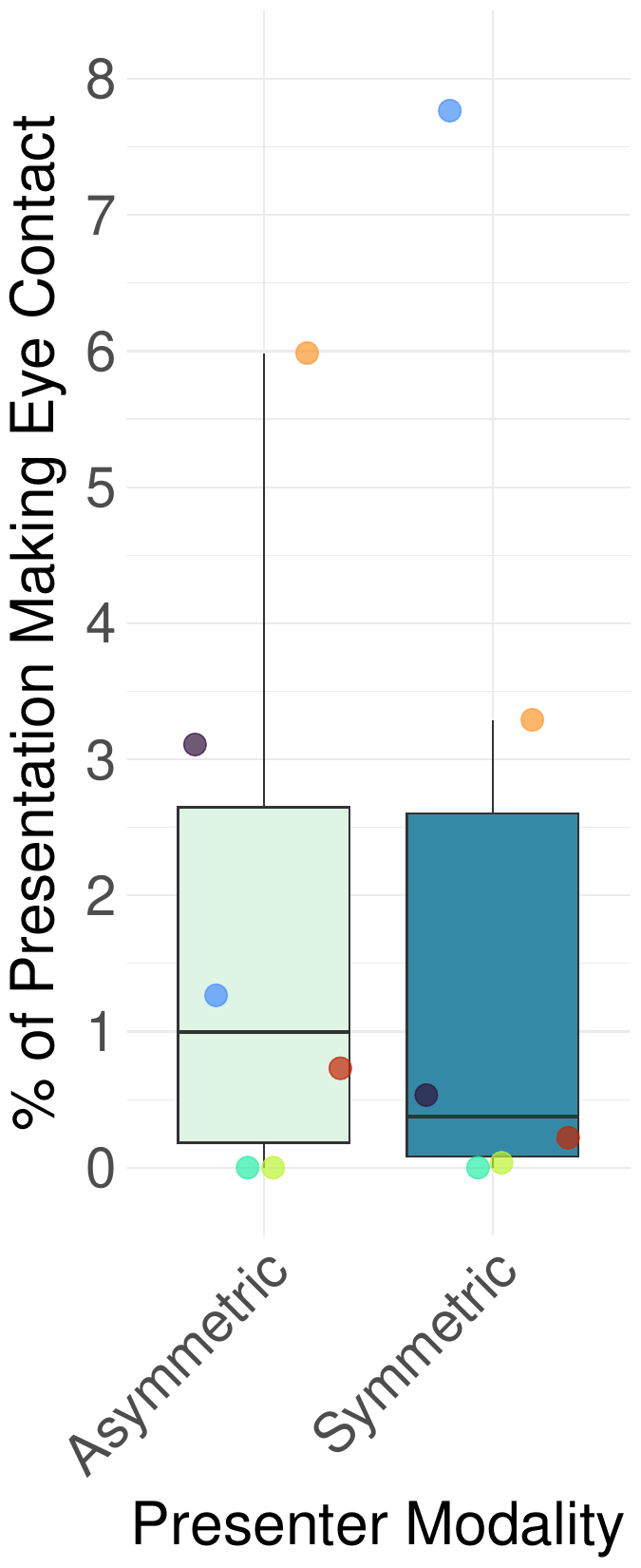}
        \caption{}
        \label{fig:eye_percent}
    \end{subfigure}
    % Description
    \Description{
    (a) Box plot with Social Expression Count on the y-axis and Presenter Modality on the x-axis. Two groups of x-axis boxes based on Presenter Modality: asymmetric and symmetric. Two boxes in each pair with different colors: Audience as yellow and Presenter as muted blue.
    (b) Box plot with Percent of Presentation Looking at Partner on the y-axis and Presenter Modality on the x-axis. Two groups of x-axis boxes based on Presenter Modality: asymmetric and symmetric. Two boxes in each pair with different colors: Audience as yellow and Presenter as muted blue.
    (c) Box plot with Eye Contact Occurrence on the y-axis and Presenter Modality on the x-axis. Two boxes on the x-axis for each Presenter Modality: asymmetric and symmetric.
    (d) Box plot with Percent of Presentation Making Eye Contact on the y-axis and Presenter Modality on the x-axis. Two boxes on the x-axis for each Presenter Modality: asymmetric and symmetric.
    }
        \caption{Box plots showing social interactions for each presentation modality. Points are colored by presentation group and presenter: within a presenter modality, points of the same color represent participants in the same presentation session; across presenter modalities, points of the same color represent the same presenter participant.
        (a) Total number of social expressions (gesturing to each other, nodding at each other, smiling at each other, and laughing) observed for each participant and each presenter modality.
        (b) Percentage of presentation time participants looked at each other.
        % Points are colored by presentation group and presenter.
        (c) Number of times participants made eye contact during the presentation.
        % Points are colored by presentation group.
        (d) Percentage of the presentation during which participants made eye contact (i.e., looked at each other simultaneously).}
    \label{fig:socialinteraction}
\end{figure*}

\section{Qualitative Results}
In this section, we describe the four themes we derived from our analysis of the interview data: Augmented Presentation Techniques, Spatial Reference and Awareness, Social Interaction and Engagement, Interface and Operation.
% \newtext{
% Where relevant to the qualitative findings, we report descriptive results of our quantitative measures such as questionnaire response scores and behavioral observation counts.
% To see tables reporting all of the quantitative results, refer to~\autoref{sec:appendix}.
% }

\subsection{Theme 1: Augmented Presentation Techniques}
This theme describes how presenter participants were able to deliver an augmented presentation experience and how audience participants enjoyed specific presentation techniques. It is divided into three categories: Immersive Data Display, Blending 2D and 3D Content, and Free-form Exploration. We found that participants, both presenters and audience, were eager to discuss the techniques they would employ when given an augmented presentation system like ours, and many participants expressed a desire to use our system for their daily work.

\subsubsection{Immersive Data Display}
Our audience participants were particularly impressed by the immersive data displays offered by the augmented presentation system.
% \newtext{The median of audience participants' scores on the presentation engagement questionnaire (see~\autoref{tab:engagement}) was $6.39$ (IQR: $0.59$, $5.97 - 6.56$).}
They appreciated the novel and engaging nature of 3D data visualizations, which gave them fresh perspectives and a deeper understanding of the data compared to plain numbers. The presence and immersion experienced in the 3D environment helped them continuously engage with the presentation content, providing a more focused audience experience. 
% need an audience quote here. AU03 immersed into data or AU11 live data is better than excel sheets.
For example, AU02 said, ``it was a really cool way to digest information, just to experience something versus having information.''
Additionally, AU03 said,
% and once I zoomed in to certain amount, now I could get even closer, and see very detailed on the heights or different counties, 
``not only can I zoom and make [the map data] bigger and then I can also be actually in the map, be actually one with the data, which I found is most intriguing. I was like in the middle of the map
% , so 3 dimensions must have made 
[...] this room [could be] filled with data. [I was] fully immersed, and you can actually experience and feel the data.''
% \newtext{The median of presenter participants' scores on the engagement questionnaire was $6.28$ (IQR: $1.28$, $5.53 - 6.81$).}
% For presenter participants, 
\changetext{Presenters noted that} the immersive data display offered them a more effective way of explaining complex narratives, making information more accessible to comprehend for the audience. 
% : uh we interact in the like A three D space which is, which can feel very like A Noble to, to the audience. So he, I think the audience is more interested and uh engaged 
PR04 said, ``[when] we interact in the 3D space...I think the audience is more interested and engaged.''
PR06 noted that the dynamic content and the interactivity were very useful for drawing the audience's interest to the presentation.

\subsubsection{Blending 2D and 3D Content}
Participants pointed out the necessity of having traditional 2D content in an augmented presentation experience, even though 3D visualization may offer a higher engagement than 2D materials. Part of the reason for blending 2D content within a 3D environment is the need to integrate existing work documents into a presentation in a new format. Some participants suggested that they would want to pull a PDF document along with a geo-spatial 3D data visualization, while others mentioned that not every type of data is good to be visualized in 3D. 
For example, AU09 said that the 3D content could complement existing slideshow materials: ``infographics or [...] textual things [are] still easier to put on PowerPoint and [you can show them] in addition to these [3D] data visualizations.''
AU03 said, ``over-utilizing the immersion can be a waste of time depending on the data,'' and mentioned that it would be beneficial to have some traditional data shown in 2D and conclude with a scene that ``brings the whole presentation into one immersive dataset.''
% quote by au03 or au09

\begin{figure*}[t!]
    \centering
    \begin{subfigure}{0.33\linewidth}
        \centering
        \includegraphics[height=21em]{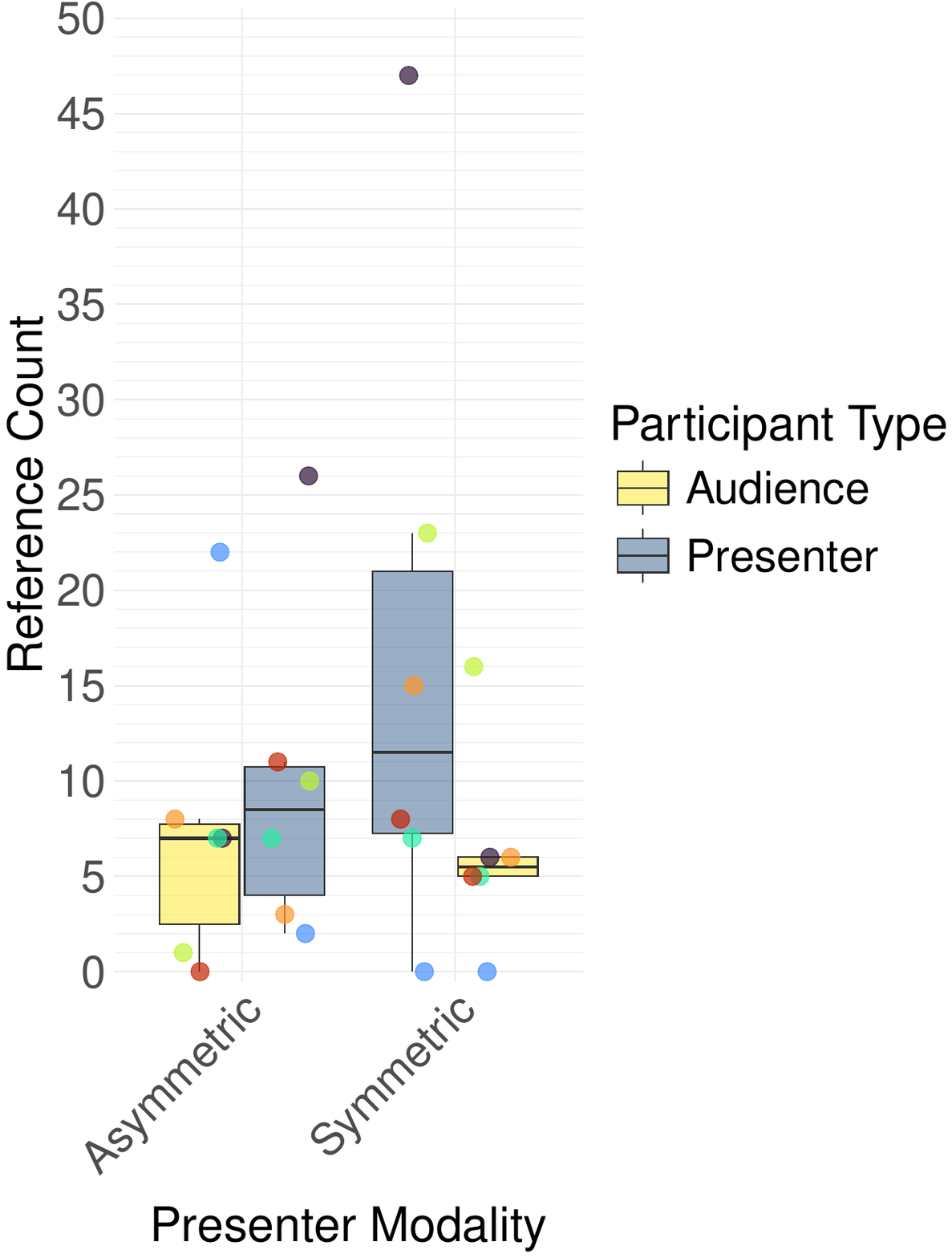}
        \caption{}
        \label{fig:ref_box}
    \end{subfigure}
    \begin{subfigure}{0.45\linewidth}
        \centering
        \includegraphics[height=18em]{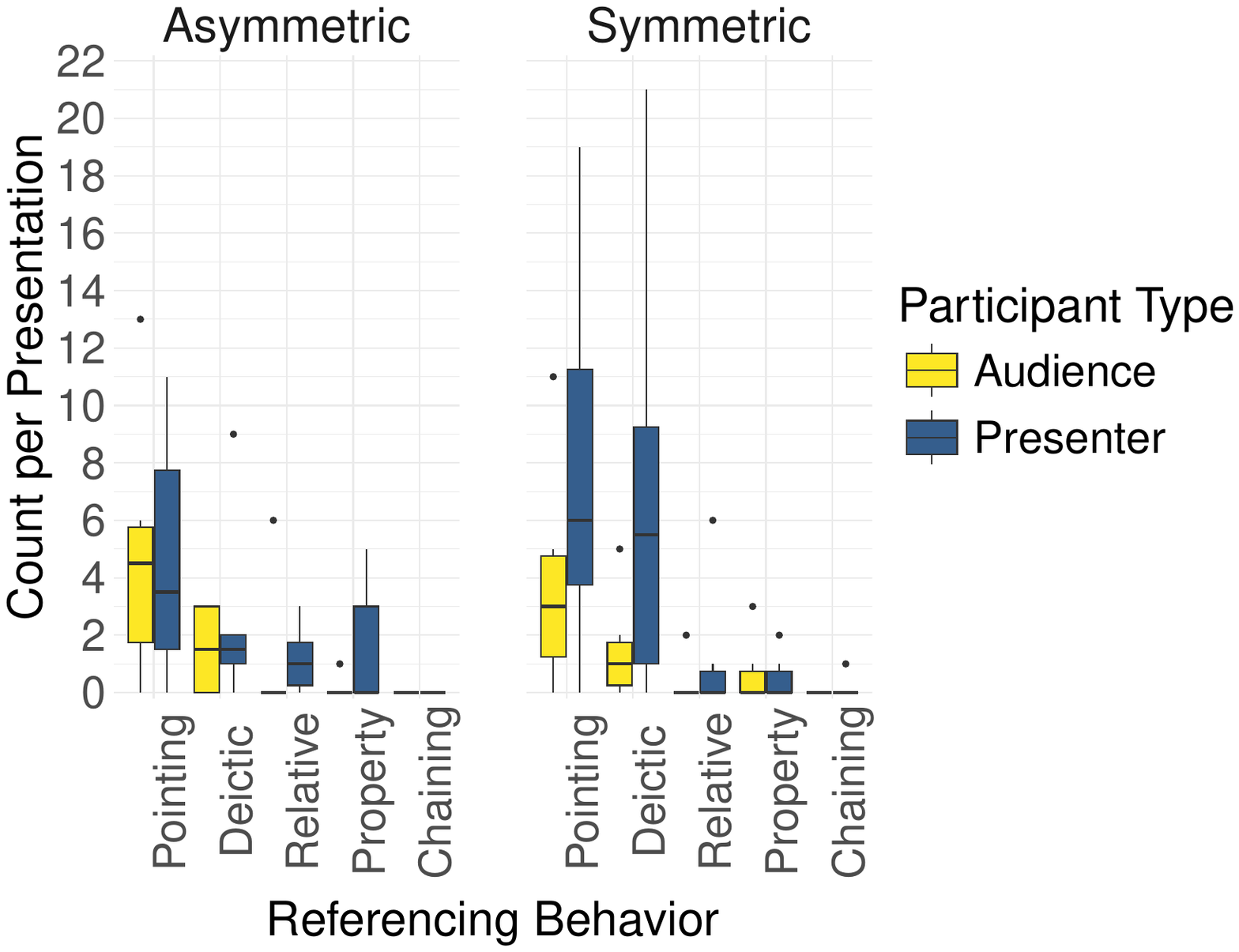}
        \caption{}
        \label{fig:ref_count}
    \end{subfigure}
    % Description
    \Description{
    (a) Box plot with Reference Count on the y-axis and Presenter Modality on the x-axis. Two groups of x-axis boxes based on Presenter Modality: asymmetric and symmetric. Two boxes in each pair with different colors: Audience as yellow and Presenter as muted blue.
    (b) Two box plots side-by-side with Count Per Presentation on the y-axis and Referencing Behavior on the x-axis. The box plot on the left is for Asymmetric presenter modality, and the plot on the right is for Symmetric presenter modality. Referencing Behavior in each plot is further broken down into Pointing, Deictic, Relative, Property, and Chaining. There are two boxes in each referencing behavior: one for Audience in yellow and one for Presenter in blue.
    }
    \caption{Participant referencing behaviors.
    (a) Box plot of the total number of references participants made during a presentation for each participant and each presenter modality.
    Points represent individual participants and are colored by presentation group and presenter: within a presenter modality, points of the same color represent participants in the same presentation session; across presenter modalities, points of the same color represent the same presenter participant.
    (b) Box plot showing the counts of referencing behaviors observed in a presentation session for each participant type and each presenter modality, including \textit{pointing} to virtual objects, using \textit{deictic} speech (e.g., ``this object''), using \textit{relative} language (e.g., ``to the left''), using a \textit{property} reference (e.g., ``the orange cylinder''), using reference \textit{chaining} (e.g., ``next to the last object referenced'').}
    \label{fig:refs}
\end{figure*}

\subsubsection{Free-form Exploration}
Our audience participants enjoyed the opportunity for interactive exploration while the presenter gave the presentation. The interactivity and flexibility allowed them to engage with the content in a more open-ended and personalized manner. 
Participants valued the freedom and control they had over their presentation experience.
% \newtext{Indeed, audience participants reported a median score of $6.00$ (IQR: $1.00$, $6.00 - 7.00$) for the \textit{Curiosity} sub-scale of the engagement questionnaire (see ~\autoref{tab:engagement}).}
Exploring the content independently while the presenter is presenting and talking in parallel enhanced the audience's sense of ownership of the presentation materials. 
AU01 said, ``the presenter could present the slides, [and] I could interact, I could see what I  wanted to see[...] Also I have more freedom to look at what I want to, instead of being controlled by the presenter.''
% PR05 also said, I am glad the audience can drive instead of me....
From the presenter's side, 
% \newtext{they reported a median score of $6.00$ (IQR: $2.00$, $5.00 - 7.00$) for the \textit{Curiosity} sub-scale of the engagement questionnaire.}
% \changetext{Moreover}, 
PR05 said, ``I'm comfortable letting them explore the 3D object. And I kind of want to stick to my script, and then in a moment I could [...] break into [interacting with the map] if I wanted to [when they ask] can you tell me more about this?''
However, this freedom does not come without a cost. Presenter participants pointed out that giving freedom to the audience would require more profound knowledge about the topic, as the questions they would get from the audience can be less predictable. Presenters also expressed concerns during their presentation that their audience looked distracted as they explored the materials independently and engaged excessively with the 3D content.

\subsection{Theme 2: Spatial Reference and Awareness}
This theme focuses on how participants utilized different spatial reference techniques while giving or taking the augmented presentation and how they tried to gauge their partner's attention. We divide them into four categories: Gesture and Pointing Reference, Verbal Reference, View Sharing and Eye Activity, and Spatial Content Layout.

\subsubsection{Gesture and Pointing Reference}
Gesture and pointing emerged as the most popular choice for object reference during the augmented presentation experience and was adopted by most presenters and audience participants.
% \changetext{
% We observed that presenters tended to make more pointing references when using the symmetric presentation modality (Mdn.: $6.00$; IQR: $7.45$, $3.75 - 11.20$) compared to the asymmetric modality (Mdn.: $3.50$; IQR: $6.25$, $1.50 - 7.75$).
% Audience members made a similar number of pointing references across both symmetric (Mdn.: $3.00$; IQR: $3.50$, $1.25 - 4.75$) and asymmetric modalities (Mdn.: $4.50$; IQR: $4.00$, $1.75 - 5.75$). 
% }
Participants frequently used virtual rays, similar to laser pointers in traditional presentations, to direct attention to specific elements in the 3D presentation environment. Participants reported that pointing is natural and easy, and most believed that their partners were well aware of their pointing gestures. AU06 mentioned, ``I used the controller as a laser pointer.''
While the use of controllers in AR for pointing was found to be straightforward, participants also noted that the accuracy of these gestures could be further improved, especially when the data points visualized in 3D are sometimes dense. Therefore, a few participants also suggested leveraging techniques like eye tracking to enhance control and accuracy further. In addition, some presenters reported that they found it more comfortable to use virtual rays on tablets as the controllers in AR were not as accurate as they expected.

\subsubsection{Verbal Reference}
While only a few participants adopted verbal communication as their primary method of referencing during presentations, they believed verbal reference to be an effective strategy, primarily when pointing at dense data points, was ineffective. 
AU06 reported that verbal referencing was effective and helped them when pointing was not working as well as expected. At the same time, verbal references were critical in scenarios when the participants felt that the AR headset limited their visual cues.
% AU03 reported that %verbal communication is necessary (especially without eye contact).... 
AU03 reported that verbal communication was necessary ``since there is no other indicator to show that I am ready [to be directed] especially since she cannot see my eyes and [...] if we don’t verbally communicate that, there is no way for her to know that.''

\subsubsection{View Sharing and Eye Activity}
Presenters often relied on the view-sharing feature provided by our system to ensure that the audience was engaged and focused on the intended aspects of the presentation. Especially when presenters were using tablets as the presenting interface, they monitored where the audience was looking mainly by the view-sharing window on the tablet. They used this live view to adjust their attention and engagement.
PR06 said, ``on the tablet I could see directly what the camera is picking up. So you'd be like, oh, their view is wandering, I need to redirect them back to the [visualization].''
On the other hand, when both presenters and audience members were wearing AR headsets, presenters tried to use more natural ways to gauge the audience's attention.
% PR04 mentioned that they were trying to track the eye activity of the audience when wearing a headset. 
PR04 said, ``I could tell what the audience is looking at by just looking at him [...] if our two worlds are calibrated, then we just look at the same thing [and it] is more natural.''
PR05 also mentioned that knowing eye activity would help them better understand the audience's interest because ``I couldn't read their face at all [behind their headset].''

\subsubsection{Spatial Content Layout}
Participants desired more structured spatial divisions within the augmented environment, such as dividing the presentation space into multiple quadrants. This organization would help presenters and the audience maintain focus and enhance awareness by clearly delineating different areas for various types of content or interaction. However, presenters also reported challenges related to the expansive nature of AR spaces. Some needed to move their heads frequently between viewing presentation content, speaker notes, and the audience's faces. This constant movement made maintaining continuous focus and awareness of the audience's attention difficult, potentially impacting their overall understanding of the partner's engagement. In contrast, the tablet interface was noted for its more integrated components, allowing presenters to switch their visual attention quickly and effortlessly between different elements. This integration provided a higher level of awareness and ease of use, enabling presenters to stay more attuned to the audience's reactions and attention without the physical strain posed by AR environments.

\subsection{Theme 3: Social Interaction and Engagement}
This theme relates to participants being able to socially interact and engage with each other during an augmented presentation experience. We have four categories for this theme: Engagement Techniques and Judgment, Social Co-presence, Eye Contact for Social Connection, and Dialogue Encouragement. % We found....

\subsubsection{Engagement Techniques and Judgment}
During the presentation experience, presenters employed various techniques to engage the audience and judge their level of focus and involvement. Presenters expressed concern that audience members seemed too focused on the immersive 3D content, often at the cost of maintaining visual contact with the presenter. To counter this, presenters frequently used verbal engagement strategies, such as asking questions, to encourage audience interaction. 
Additionally, some presenters used 3D content manipulation, such as panning and zooming in and out on maps, to guide the audience’s attention and maintain engagement throughout different presentation sections.
% \newtext{Across both modalities, presenters who reported having weekly experience giving presentations tended to rate their ability to engage the audience more highly with a median score of $6.81$ (IQR: $0.17$, $6.72 - 6.89$) compared to $5.61$ (IQR: $0.94$, $5.25 - 6.19$) for those with occasional presentation experience.}

To gauge audience involvement, presenters observed how the audience interacted with the 3D content and noted the types of questions asked during the presentation. 
% Presenter PR06 particularly mentioned that it was \textit{``easier to tell audience engagement in tablet''} for the simplicity and better integration of the interface, making it easier to focus on observing their partner.
PR06 relied on the live audience view in the tablet for this purpose and said, ``it was easier to tell [the audience's engagement level] on the tablet because I could see their like vision moving around as like things were changing.''
PR03 mentioned that the familiarity of the tablet interface enabled them to engage the audience better: ``Because I was more comfortable with the tablet, I think I was a lot more engaging in that presentation. I didn't really have to think too much about what I was doing.''
On the other hand, most of the audience appreciated the presenters' attempt to engage them with verbal cues and thought that having someone talk to them and walk them through the presentation materials by voice helped them better understand and experience them.
%[Quote AU07 or AU08]. 
For example, AU08 said, ``Just having a voice-over to navigate the experience was really good. That definitely made it engaging.''
AU04 mentioned the need for more body and visual cues by virtual avatars, with suggestions for stronger social connections to help bridge the engagement gap.

\subsubsection{Social Co-presence}
Participants using AR headsets reported a stronger sense of intimacy and connection, with interaction styles—such as gestures and face-to-face conversations—feeling more natural than those using tablets. 
% \newtext{
% For the Interpersonal sub-scale of the presentation engagement questionnaire (see~\autoref{tab:engagement}), presenters reported approximately the same scores in both symmetric and asymmetric presentation modalities (both with Mdn.: $6.50$; IQR: $1.00$, $6.00 - 7.00$), but audience participants reported slightly higher interpersonal engagement in the symmetric modality (Mdn.: $6.50$; IQR: $1.00$, $6.00 - 7.00$) compared to the asymmetric modality (Mdn.: $5.50$; IQR: $1.00$, $5.00 - 6.00$).
% }
The symmetric setup, where both the presenter and audience used headsets, made participants feel as though they were together, experiencing the presentation content simultaneously.
% quote from au02 that this setup gave the impressi on that everyone was viewing the exact same content, Promoting a sense of equal status and shared experience. 
For example, AU02 noted, ``it made me feel like we were more kind of experiencing something together, 'cause we had the same equipment on versus just kind of the presenter just being like a separate person over there just showing the slides. It did make me feel like we are experiencing together.''
In contrast, the asymmetric setup positioned the presenter as more of a solo guide. Tablets gave the presenter a greater sense of anonymity and less direct control or involvement with the virtual content in the space.
AU07 said, ``from my perspective, I thought I was in my own environment [...] I didn't really see him in the world. I just thought of him as a voice over until started seeing the [virtual pointing ray].''
Some presenters also expressed uncertainty about whether the audience was fully engaged with the presentation or simply exploring independently, leading to concerns that the audience members might have been disconnected and immersed in their world.
For example, PR05 said, ``It made me excited that [the audience] could do their own exploration. And I was like merely their guide rather than like presenting it to them [...] but I couldn't pick up on many cues as to whether they might have been disinterested or distracted. I can't see their face.''

% \newtext{
% Despite this spatial disconnect, both presenter and audience participants exhibited slightly more social expressions (in the form of participants gesturing to each other, nodding at each other, smiling at each other, and laughing) in the asymmetric modality compared to the symmetric modality.
% }
% \changetext{
% In the asymmetric presentation modality, the median number of social expressions exhibited by the presenter was $1.50$ (IQR: $4.50$, $0.00 - 4.50$), and the audience's median number of social expressions was $4.50$ (IQR: $13.50$, $0.00 - 13.5$).
% For the symmetric presentation modality, the median number of social expressions exhibited by the presenter was $0.50$ (IQR: $2.50$, $0.00 - 2.50$), and the median number of social expressions exhibited by the audience was $2.50$ (IQR: $1.75$, $1.25 - 3.00$).
% The number of social expressions observed per participant is shown in~\autoref{fig:socialexpr_pid}.
% }

\subsubsection{Eye Contact for Social Connection}

In the interviews, participants consistently highlighted the importance of eye contact for maintaining social connection during AR presentations, but many encountered challenges due to the headset's limitations.
AU03 mentioned that they tried to make eye contact with their partner but stopped, feeling that their partner likely couldn't see their eyes, leading to a sense of uni-directional interaction. Others reported that the headset restricted their peripheral view, while some had to step back to see the entire desk or map, further complicating attempts at social engagement. AU11 and PR03 noted that ``the headset was a bit of burden'', making it difficult for the audience to rotate their heads or look at their partners naturally. In the asymmetric setup, presenters using tablets were better able to observe the audience's body language, and it was easier to socially interact with the audience when using the tablet, as ``the headset feels like a physical blocker,'' hindering direct engagement compared to the tablet.
AU03 noted, ``oh she can't see my eyes now. I can tell when she was looking at me, she was a good presenter. And [...] she knows that I am looking at her now, but it's a weird masked feeling where you know she is not able to tell if I have my eyes closed, and she would have no idea.''
% \newtext{Combined with feelings of being in separate worlds described in the previous section, these headset issues may help explain the lower interpersonal engagement score that the audience reported for the asymmetric modality.}

\subsubsection{Dialogue Encouragement}
The participants reported that the augmented presentation offered greater visual and social dynamism, leading to improved engagement between the audience and presenter and encouraging more interaction and Q\&A. Participants, such as AU01 and PR05, observed that the enhanced interactivity naturally prompted more feedback and questions from the audience. 
% For example, PR05 noted a time when they were interacting with the audience and it led him to improvise off-script.
However, both presenters and audience members noted that this type of presentation demands more time for discussion, emphasizing that presenters need to be well-prepared for unpredictable questions. 
For example, PR05 said, ``In an ideal scenario, I would go in the immersive environment with [the audience] and we'd have a conversation about [the presentation content]. I'd be knowledgeable enough about it that we would just have a conversation.''
Additionally, AU01 said, ``the presenter has to be more creative in giving the narrative [...] because the [audience] can do anything [...] You have to be more knowledgeable about what you are going to present because there will be more types of questions that come out.''
This requires a higher familiarity with the material than traditional PowerPoint presentations. Participants felt that the interactivity of augmented presentations fosters more conversation, especially for small focus groups, with AU05 adding that whereas in ``some mundane [presentations], people start to fall asleep,'' but the immersive presentation ``keeps you up.'' 
% one participant adding that the audience would ``never fall asleep.''

Some presenters also commented on the differences in the interface. The tablet was generally seen as a familiar and easy-to-use tool with less overhead for the presenters that encouraged social interactivity. For example, one presenter felt that the headset didn't provide the level of control they expected, making them concerned about unintentionally ``jarring'' the audience. Only one presenter expressed that the headset allowed more natural conversation as the tablet does not allow enough space for gesture cues.

\subsection{Theme 4: Interface and Operation}
This theme presents participants' feedback on the interface impact of AR and tablet devices and how different interfaces affected their operation while conducting the presentation. This theme has three categories: Control Competition, Perception of Interfaces, and Technology Adaptation. % We found...

\subsubsection{Control Competition}
Participants reported managing shared control and ownership over the 3D map between the presenter and the audience. Several participants reported issues with competing for power, resulting in jumpy and disjointed map manipulations. PR01 specifically pointed out that this competition hindered the fluidity of the presentation. To avoid such conflicts, some audience members opted to avoid directly interacting with the map, instead pointing to areas of interest to guide their partners subtly. Others suggested more structured solutions, such as implementing a turn-based control system, allowing the audience to take some extra time to freely explore the content after each slide, especially since the immersive and interactive data display requires more time to comprehend, as mentioned by AU03: ``every time the data changed, it took time to talk through it [...] If you had a PowerPoint presentation, that would be one thing, but if you change the map and the landscape of the environment you are in, it's like you are having to [...] take back in the changed environment.'' Some participants, including AU01 and PR01, proposed giving the audience a dedicated window or interface for controlling the map to reduce competition. PR04 also expressed the need for the ability to reset the map after the audience had manipulated it to maintain the presenter's control over the presentation flow.

\subsubsection{Perception of Interfaces}
Participants reported that different setups could vary the presenter’s and audience's experience during an augmented presentation.
% \changetext{
% On the NASA-TLX cognitive load questionnaire, presenters reported a slightly higher mental demand score for the symmetric presentation modality (Mdn.: $3.50$; IQR: $2.50$, $2.25 - 4.75$) compared to the asymmetric modality (Mdn.: $3.00$; IQR: $2.25$, $1.50 - 3.75$).
% Presenters reported similar physical demand scores for both the symmetric modality (Mdn.: $2.50$; IQR: $2.50$, $1.25 - 3.75$) and asymmetric modality (Mdn.: $2.50$; IQR: $1.75$, $1.25 - 3.00$).
% Presenters reported similar performance scores for the symmetric modality (Mdn.: $5.50$; IQR: $0.75$, $5.25 - 6.00$) and the asymmetric modality (Mdn.: $5.50$; IQR: $1.75$, $5.00 - 6.75$).
% }
Three presenters preferred the tablet because it provided a familiar interface, a better overview, and easier presentation control.
% with PR01 and PR06 mentioning specifically that the tablet ``allowed for a clearer vision of the content'' while speaker notes were reported to be difficult to read while using the headset. 
For example, PR06 said that with the tablet, ``it was easier to keep track of everything'' whereas the headset interface was ``a bit distracting''. Furthermore, R03 and PR06 noted that positioning the presentation controls in the headset interface caused additional overhead.
At the same time, the headset was perceived as fostering a more collaborative and natural interaction due to its symmetric nature, where both the presenter and the audience engage with the content equally. The tablet, on the other hand, reinforced the role of the presenter, with participants feeling more in control and in a traditional presentation mode while holding it. AU03 added a perspective on the context of co-located interactions, expressing a preference and necessity for maintaining a connection via the physical environment. They argued that just as it doesn't make sense to have Zoom meetings when both users are physically present; using excessive immersive space in situations where users are co-located also can make people feel unnecessary.

\subsubsection{Technology Adaptation}
Our participants expressed strong interest in using the system for their daily work, but they also highlighted several factors necessary for smoother adoption. One key aspect was the need for an easy-to-use authoring method, as creating 3D visualizations for presentations requires both time and technical skills. Both presenters and audience members mentioned the importance of having simple tools to quickly build 3D presentations and easily map and convert 2D data into 3D formats. Participants also emphasized the value of standardized design patterns, templates, or frameworks for 3D scene slides that anyone could use readily, reducing the barrier to preparing such an experience.

Some participants also reported that working in 3D was cognitively demanding. Presenters, in particular, found it challenging to manage speaker notes, manipulate 3D content, and engage the audience simultaneously. 
PR05 noted, ``Being immersed was overwhelming, so it was harder to focus on my story and my speaker notes.'' 
Participants also wanted physical interaction within the AR headset to feel more natural and for the device to be more comfortable. The frequent head movement required in an immersive mode made them desire lighter headsets for better comfort. A few participants mentioned feeling dizzy or experiencing headaches during the presentation, underscoring the importance of improved hardware for better user adoption and overall experience.

\section{Discussion}

This section presents design guidelines for creating immersive presentation systems derived from our main findings.
Additionally, we outline the limitations of the present study.
Lastly, we present ideas for future work to build on our findings.

\subsection{Design Guidelines for Immersive Presentation Systems}

% Based on participants' questionnaire responses and interview data, we found that immersive technologies can provide an engaging presentation experience.
% for presentations of data-driven content.
\changetext{This section offers design guidelines for researchers and practitioners aiming to create \newtext{co-located} immersive presentation systems that maximize user engagement, with insights on selecting suitable presentation formats and key considerations for content preparation to ensure meaningful and interactive experiences.}

\subsubsection{Support balanced interactivity and audience exploration}
% Use data that is well-suited for display in 3D.
% Give the audience an opportunity to explore something on their own.
% Allow collaboration, but make ownership of interactive content clear.
% Interactivity encourages dialogue.
In the interviews, participants reported feeling engaged with the novel presentation and being particularly drawn to the content's immersiveness and interactivity.
They also noted how the 3D content complements more traditional 2D presentation content.
To best support interactivity in augmented presentations, it is important to use data visualization that is well-suited for 3D display, as there are significant trade-offs between the benefits of 3D interactivity and the challenges it presents~\cite{Isenberg2018immersive, lee2021visceral, saffoIASpace}. On the positive side, 3D interactivity allows the audience to explore the content independently, promoting a deeper understanding of the subject matter~\cite{lee2021visceral} and potentially encouraging more conversation and Q\&A, as observed in our study. However, as our participants noted, this increased audience engagement also places more demands on the presenter, who must be highly familiar with the details of the dataset. The presenter needs to be prepared for unpredictable audience exploration~\cite{lee2021visceral} and questions or observations from the audience, and potentially some difficulty controlling the pace of the presentation.
% making it more difficult to control the pace of the presentation.
% Moreover, when the audience is given more interactivity, the presentation can shift toward 
This can shift the presentation toward a more collaborative experience, with the audience influencing how the content is displayed or introducing new topics based on their discoveries. 
\newtext{As indicated by our qualitative data results, this sense of collaboration is especially stronger in a symmetric setup, characterized by a sense of equality, which fosters a stronger role for the audience in terms of their control capability. We also observe a similar trend in the quantitative data, suggesting that a symmetric setup could potentially lead to better group awareness. This aligns with the findings by Drey et al. that a symmetric setup improves communication and workspace awareness in paired VR learning \cite{drey2022collaborative}. 
However, our presentation scenario is a more uni-directional and structured experience compared to paired learning.
% a major difference between presentation and learning is that it is frequently a more uni-directional and structured experience.
} 
% \newtext{As indicated by both our quantitative and qualitative data results, this sense of collaboration is especially stronger in a symmetric setup, characterized by a higher level of group awareness and sense of equality, fostering a stronger role for the audience regarding their control capability.} 
The shift from guided presentation towards collaboration can reduce the presenter's control and ownership over the flow of the presentation. This level of audience interactivity needs to be carefully considered depending on the style and purpose of the presentation. \newtext{Different from Drey et al.'s guideline that the experience should always allow both users to move and interact freely in the VE and that symmetrical setups is always superior, our work highlights the unique trade-offs between symmetric and asymmetric settings in augmented presentation.}
\changetext{\textbf{We recommend balancing interactivity with the presenter's ability to guide and manage the overall experience based on the presentation goals, recognizing that symmetric setups can potentially facilitate easy collaboration with the audience, whereas asymmetric setups offer the presenter good control and direction.}}
% , ensuring that the format aligns with the intended goals of the presentation.

\subsubsection{Support social interaction between the presenter and audience}
% \changetext{
% In related research examining paired learning in co-located settings, Drey et al. found that their symmetric condition improved communication and workspace awareness, leading them to suggest using symmetric systems whenever possible.
% They also suggested that in asymmetric VR setups, the tablet user should be able to freely explore and interact in the virtual environment to improve their spatial awareness.
% Furthermore, they suggested providing avatars for users and correcting any physical-virtual mismatch between audio or visual cues that may result from virtual locomotion.
% Last, they recommended explicit signaling within the VE when guiding VR users to look at specific objects.
% In addition to workspace awareness, our study is particularly interested in the effects of different system configurations on users' social interactions.
% }

In observing participants' behaviors during the presentations, there were more social interaction occurrences (e.g., nodding or gesturing to each other, smiling) when presenters used the asymmetric presenter modality compared to the symmetric modality.
\changetext{Both presenters and audience} also made more eye contact with greater frequency and for longer durations when using the asymmetric presentation modality.
During interviews, participants noted that the reduced observable social cues caused by wearing the HWD led to lower-quality social interaction.
This finding aligns with related research suggesting that XR headsets may limit typical social interactions because they occlude a large portion of the wearer's face, therefore preventing natural eye contact and complete observation of facial expressions~\cite{mcatamney2006examination, viola2022seeing}.

On the other hand, we observed that audience members tended to spend more time looking at the presenter when they used the symmetric modality.
Additionally, in interviews, audience members reported feeling like they were in a separate world from the presenter when they used the asymmetric presentation modality because the presenter was not immersed in the same AR space.
In the symmetric condition, participants reported the opposite: they felt part of the same experience.
While we did not ask participants about their perceptions of co-presence~\cite{harms2004internal}, future work should investigate this presentation scenario further in this regard.

It is worth considering design compromises to resolve the tension between reduced social interactions and increased time spent looking at the presenter in the symmetric scenario.
One solution explored in the literature is to display an HWD user's eyes on the outside of the display to restore eye gaze cues and awareness and increase social presence with others~\cite{bozgeyikli2024googly, chan2017frontface, mai2017transparent, matsuda2021reverse}.
\newtext{The development of photorealistic avatars \cite{kartik2024, chen2024avatar} and enhanced facial animation generation \cite{zhao2024m2f, sun2024diffposetalk} in addition to Apple's Personas\footnote{\url{https://www.apple.com/apple-vision-pro/}} and Meta's codec avatars\footnote{\url{https://about.meta.com/realitylabs/codecavatars/}} 
% (e.g., Apple's Personas\footnote{\url{https://www.apple.com/apple-vision-pro/}} or Meta's codec avatars\footnote{\url{https://about.meta.com/realitylabs/codecavatars/}} 
could be used to overlay an avatar on users with HWDs to preserve observation of their social signals.}
From the presenter's side, \newtext{when using the tablet interface} it may be useful to include an AR view so that the presenter may peer into and interact with the AR space directly.
% Additionally, displaying visual effects, such as virtual hands or annotation rays, that better suggest how the presenter's real-world movements affect the virtual content can strengthen feelings of co-presence~\cite{kim2018social}.
Another option is for the audience to use a tablet with an AR view, which will place both the presenter and audience in the same world and reduce social interaction blockers for both users.
However, we recommend that the audience use the AR HWD due to its advantages in permitting more direct interaction and increased immersiveness.

\textbf{%Overall, 
In cases that require emphasizing social connection between the presenter and audience through typical social behaviors, we recommend reducing equipment that inhibits the presenter's social expressions, e.g., by using a tablet to control the presentation.
% In cases where 
When it is more important that the presentation feel like a shared experience, we recommend that both the presenter and audience use AR HWDs.}
% \note{MG: Added these last two sentences, see if it works. When Fannie read this section she commented ``Seems like social cues matter for both presenter and audience, so I wonder why the recommendation isn't to reduce equipment on both sides, i.e., if you have a AR view for presenter on the tablet you could have a tablet AR view for the audience too. they would also be able to feel like they're in the same world with the benefits of a symmetric modality.''}

\subsubsection{Support intuitive referencing and awareness}
As noted in our observations of participant behavior and their interviews, participants made spatial references to virtual content using both pointing and verbal communication during the presentation to direct their partner's attention.
As shown in~\autoref{fig:ref_box}, presenters tended to make more references when using the symmetric modality.
More specifically, as shown in~\autoref{fig:ref_count}, when using the symmetric presentation modality, presenters pointed and used deictic phrases to refer to content more frequently than in the asymmetric modality.
Additionally, as shown in~\autoref{fig:awareness}, audience members appeared to experience greater group awareness with the presenter (e.g., they could better understand each other's references and attention) in the symmetric condition.
In the interviews, participants agreed that referencing was intuitive in the symmetric modality. Still, the AR HWD blocking gaze direction prevented complete awareness of where their partner's attention was and what they were interested in.
This point is another reason to consider displaying the HWD user's eyes outside of the headset~\cite{bozgeyikli2024googly, chan2017frontface, mai2017transparent, matsuda2021reverse} or adding visualizations to represent a user's orientation and gaze~\cite{erickson2020sharing, piumsomboon2019collab}.

Some participants reported successfully using the tablet's view sharing and pointing ray features to make references, which aligns with related research employing similar referencing techniques in asymmetric collaboration scenarios~\cite{johnson2021referencing, schroeder2023dyadic}.
However, the lower scores for the group awareness questionnaire, the fewer references made, and participants' comments about feeling as though they are in separate worlds for the asymmetric modality suggest these features are not equivalent to a symmetric experience in providing a common ground for spatial referencing. \newtext{This highlights a unique challenge in AR compared to the VR setting studied by Drey et al. \cite{drey2022collaborative} --- the prominent presence of the physical environment in AR for both co-located parties necessitates more aligned levels of awareness and referential cues, for which the less direct signaling approaches could fall short.}
One participant suggested the tablet could have an AR view option to address this issue.
\textbf{We recommend providing view and interaction symmetry (e.g., by both presenter and audience using AR HWDs or by providing an AR view through the presenter's tablet) in cases where group awareness is critical, such as when the presentation calls for lots of shared attention and interaction around the presentation content.}

% In the interviews, participants reported that being able to judge their partner's gaze is important for understanding what they are attending to and interested in.

\subsubsection{Consider interface familiarity and ergonomics}
% It is important to give people something familiar and seamless to incorporate in their normal lives.
% Some people will also be able to take on a larger cost to learn things if there is a big value added.

Immersive technology offers a much larger space to display content and provides a more flexible presentation control interface, allowing for greater expressiveness in presentations. However, this increased space and flexibility can introduce challenges, such as interaction and visual fatigue, disorientation, and higher cognitive and physical effort. 
% When virtual windows and panels become too sparse or distant, especially if they are easily movable by the user, it can often lead to a less organized workspace. 
% We found that participants reported lower mental demand for the tablet interface, and in the interviews, some reported having difficulty interacting with the AR HWD interface.
\changetext{Our quantitative data suggests a trend toward lower mental demand when using the tablet interface, and in the interviews, some participants reported having difficulty interacting with the AR HWD interface.}
While presenters ideally would adjust the interface to suit their ergonomic needs, in practice, presenters are often under heavy cognitive load during a presentation. They have to multitask for a presentation, such as engaging the audience, delivering expressive narration, and monitoring audience behavior, making it difficult to focus on arranging an ergonomic interface layout based on the changing position of the audience.

A key finding from the interviews was that three of the presenters preferred using the tablet over the AR HWD because it provided better integration of content and control, allowing them to manage everything --- control the slides, observe the content and the audience's focus, read scripts, and engage with the audience --- within one centralized space, requiring far less physical movement. Presenters also commented that the familiarity of the tablet interface with other presentation tools they already use made it easier to deliver a good presentation. This familiarity integration becomes even more critical in real-world workplace settings, where presenters must deliver multiple presentations in succession.
\textbf{We recommend using highly integrated and conventional interfaces to minimize
% physical and 
mental load and make the presentation system more practical for sustained use, particularly in professional environments where efficiency and endurance are vital.}
In contrast, a sparse and disjointed interface layout can quickly become labor-intensive and lead to faster fatigue, undermining the benefits of immersive presentation technology.

\subsubsection{Pre-presentation Preparation and Real-time Content Manipulation}

When designing authoring systems for augmented presentations, it is important to consider both pre-presentation preparation and real-time manipulation during the presentation itself. Pre-presentation authoring should resemble the process of preparing traditional slides, where the presenter maintains complete control over the content, structure, and flow ahead of time. 
% This method allows careful planning, giving the presenter time to organize information and 
This method helps to ensure a smooth delivery. The advantage is that the presenter can focus on delivering the content without manipulating the data or visuals in real-time, reducing cognitive load during the presentation.

In the interviews, our participants highlighted that having tools and design templates for easy creation of augmented presentations, such as 2D to 3D data mapping tools and PowerPoint slide conversion into 3D formats, would greatly ease the onboarding process. 
% Strong compatibility with familiar platforms and tools to scaffold the experience would reduce the learning curve and provide users with a familiar starting point for creating 3D presentation content. 
This approach would allow us to build on established user-centered design principles from web and desktop interfaces~\cite{chenm2021}. \newtext{At the same time, our participants indicated a preference for using 2D and 3D materials for different situations and purposes. Traditional documents like PDFs and existing 2D slides can serve well without introducing additional preparation overhead, while 3D visual content is ideal for interactive data exploration and sense-making.} \changetext{Especially when it comes to 3D content experience, real-time manipulation of presentation content allows the presenter to tailor the experience to the specific needs and interests of the audience. As we derived from the interviews, this dynamic approach can enhance audience engagement by making the presentation more interactive and personalized.} However, it also introduces a higher cognitive burden for the presenter, as they must manage both the content and its presentation on the fly while engaging the audience. This type of authoring requires a balance between flexibility and control, ensuring that while the presenter can make real-time adjustments, the interface and tools do not overwhelm them with complexity. 
\textbf{We recommend designing a system that supports intuitive pre-planning and real-time adjustments, \changetext{effectively integrating 2D and 3D content based on actual needs of interactivity to enhance the effectiveness and versatility of augmented presentations. By doing so, presenters can leverage the strengths of both content types to deliver more compelling and contextually relevant presentations while minimizing friction during the preparation process.}}

\subsection{Limitations}

One limitation of our exploratory user study is that it involved a small sample size (n = 12 groups).
Our data showed variance across groups, potentially due to individual participant differences (e.g., presenters who minimally engaged their audiences).
While our results still provided valuable insights, the limited number of participants may impact the generalizability of the findings to the broader population. 
% Future research should aim to recruit a larger, more diverse sample to validate and extend the results of this study.

In real-world presentation scenarios, presenters are often familiar with their presentation material before the presentation.
However, in our user study, participants \newtext{were not experts on the subject matter} and had limited time to familiarize themselves with the presenter interfaces and practice the presentation, \newtext{which limits the generalizability of our findings.}
\newtext{While no participants complained about having enough time to practice,} this could have reduced presenters' confidence in the system and the presentation content, which could have also reduced their performance or social interactions.
A lack of confidence in the system or content might explain why some presenters hardly interacted with their audience during the presentation and focused mainly on advancing the slides and communicating the messages on the speaker notes.

% Our analysis of participants' social interaction when using the presentation system should be considered in light of the fact that some participants hardly interacted with each other.
% While we instructed presenter participants to try to engage the audience interpersonally as well as with the presentation content itself, some presenters and audience members exhibited no social interaction behaviors during the presentation (see ~\autoref{fig:socialinteraction}).
% This lack may be explained by several factors.
% First, participants may have different presentation styles and choose to engage their audience in different ways.
% Additionally, participants were not familiar with the presentation before the experiment and had limited time to practice the presentation, which could have reduced their comfort with the material and led to lower overall confidence.
% Similarly, learning a new interface to give the presentation may have reduced participants' confidence in their presentation.
% Last, participants did not know each other before the experiment and may not have felt as socially comfortable around each other.
% Future work on interactive augmented presentations could more intentionally insert interaction with the audience in the presentation script to try to elicit social behaviors from all participants.

\subsection{Future Work}

To address the lack of social behaviors exhibited by some presenters, one avenue for future work is to provide automatic and explicit cues to the presenter to engage the audience, e.g., by periodically reminding them to interact socially with the audience.
Additionally, a system could detect the audience's engagement level throughout the presentation (e.g.,~\cite{hassib2017engage}) and automatically provide feedback to the presenter about which kinds of content or interactions are most engaging for the audience.

Likewise, we observed that the ability of the audience to explore the data on their own was seen as a benefit by both audiences and presenters, with the latter sometimes encouraging this exploration. However, there will still be times when the presenter demands the audience's full attention, such as when introducing an important concept. Thus, proper mechanisms in facilitating transitions between controlled and exploration presentation phases may be investigated in the future, whether it is to provide explicit (or implicit) cues for when the audience is expected to explore the data or even to devise appropriate techniques to ``gracefully interrupt'' an audience member who may be engrossed in the data.

While we focus primarily on an intimate tabletop setting for this work, there are myriad more scenarios and contexts where presentations are often conducted, which can benefit from AR.
The number of audience members can easily vary, ranging from small groups (e.g., a family) to large crowds (e.g., a conference talk), and there can be more than one presenter, be it multiple presenters swapping once throughout the presentation or them interweaving their narration between each other. These simple differences introduce new challenges, for example, how larger groups of people share the physical space during co-located augmented presentations, how presenter(s) keep awareness of what their audience member(s) are looking at, and how the audience member(s) might want to explore the data for themselves. 
\newtext{Future work could also explore how using other virtual augmentations such as visualizing the AR HWD user's eyes on the outside of the headset (e.g., as studied in research~\cite{bozgeyikli2024googly, chan2017frontface, mai2017transparent, matsuda2021reverse} and implemented in the Apple Vision Pro) or using VR in co-located presentations affect users' experiences.}

\newtext{Additionally, our study results may be influenced by novelty effects, as some participants were new to AR technology. A future longitudinal study could help us better understand how users' social interactions and engagement might evolve as they become more familiar with the system. Such a study could also involve usage in realistic scenarios, where participants present their own materials, allowing for a more comprehensive evaluation of the system's usability and effectiveness over time and in real-world conditions.}

The space that the augmented presentation is occurring in can also vary, mainly when the presentation is situated around some physical context~\cite{willett2017embedded}. For example, a presentation for prospective home buyers might involve walking around different parts of the house, or a presentation for car dealerships might involve buyers physically inspecting multiple cars side by side, guided by a presenter. The need to navigate and coordinate presentation content in large spaces poses new challenges, for example, how the presenter might be aware of the location of presentation content without (or even with) an AR HWD, and how they might ensure proper attention guidance and engagement for audiences who may become distracted in a more unconstrained space.

% encouage better presentation styles explicitly.. encourage audience interaction etc.
\section{Conclusion}
Motivated by the use of AR to facilitate immersive and engaging presentations to audiences, we developed an augmented presentation prototype system and investigated how the choice of presenter modality---immersive AR (i.e., symmetric) versus non-immersive tablet (i.e., asymmetric)---influences engagement, group awareness, and social interaction between the presenter and audience.
We conducted a user study of 12 participant pairs wherein a presenter, either with an AR HWD or a tablet, delivered a presentation to an audience member wearing an AR HWD.
% Our results show that participants found the presentation system engaging and were particularly drawn to its immersiveness and interactivity.
% We also found that participants performed more social interactions when using asymmetric modalities, but audiences spent more time looking at the presenter with symmetric modalities. Presenters in symmetric modalities also relied more on pointing gestures and deictic phrases to point out areas of interest on the presentation content.
Based on participants' experiences, we derived four themes around presentation techniques, spatial reference, social interaction, and interface impact.
We provide design guidelines for immersive presentation systems based on these findings and propose several new directions to support future research and real-world applications of augmented presentations.
% \newpage


% \begin{acks}
% \end{acks}

\bibliographystyle{ACM-Reference-Format}
\bibliography{sample-base}

% \newpage
\appendix
\onecolumn
\section{Supplementary Materials}
\label{sec:appendix}

This supplementary material provides tables showing the specific questionnaire questions and more fine-grained results of participants' questionnaire responses and the behaviors we observed.

\subsection{Questionnaires}

\begin{table*}[h]
    \begin{tabular}{ | m{4em} | p{7em} | p{30em} | } 
        \hline
        \textbf{Category} & \textbf{Element} & \textbf{Question} \\ 
        \hline
        What & Action & I always knew what my partner was doing. \\ 
        & Intention & I always understood the goal of my partner’s actions. \\ 
        & Artifact - Partner & When they referred to something, I always knew what object my partner was referring to.  \\ 
        & Artifact - Self & My partner could always understand what I was referring to. \\ [1ex]
        Where & Location & I always knew where my partner was located. \\ 
        & Gaze & I always knew where my partner was looking.  \\ 
        & View & I always knew what my partner could see.  \\ 
        & Reach & I always knew where my partner could reach.\\ 
        \hline
    \end{tabular}
    \caption{Group awareness questionnaire.}
    \label{tab:awareness}
\end{table*}

\begin{table*}[h]
    \begin{tabular}{ | p{7em} | p{35em} | } 
        \hline
        \textbf{Category} & \textbf{Question (presenter adaptation)} \\ 
        \hline
        Attention Focus & This presentation medium keeps me (my audience) totally absorbed in the presentation. \\ 
         & This presentation medium holds my (audience's) attention. \\ [1ex]
        Curiosity & This presentation medium excites my (audience's) curiosity. \\ 
         & This presentation medium arouses my (audience's) imagination. \\ [1ex]
        Intrinsic Interest & This presentation medium is fun (for my audience). \\ 
         & This presentation medium is intrinsically interesting (for my audience). \\ [1ex]
        Overall & This presentation medium is engaging (for my audience). \\ [1ex]
        Content & I was engaged (was able to engage my audience) with the presentation’s virtual content. \\ [1ex]
        \changetext{Interpersonal} & I was engaged with the presenter themselves. (I was able to engage my audience interpersonally/socially.) \\ 
        \hline
    \end{tabular}
    \caption{Presentation engagement questionnaire.}
    \label{tab:engagement}
\end{table*}

\newpage
\subsection{Questionnaire Responses}
In these tables, IQR indicates the interquartile range of the data. Q1 and Q3 represent the first quartile (25\%) and the third quartile (75\%) of the data, respectively.

\begin{table*}[h!]
\begin{tabular}{|l|l|l|l|l|l|}
\hline
\textbf{Subscale} & \textbf{Modality} & \textbf{Median} & \textbf{IQR} & \textbf{Q1} & \textbf{Q3} \\ \hline
\multirow{2}{*}{Mental Demand}   & Symmetric  & 3.50 & 2.50 & 2.25 & 4.75  \\ \cline{2-6} 
                                 & Asymmetric & 3.00 & 2.25 & 1.50 & 3.75  \\ \hline
\multirow{2}{*}{Physical Demand} & Symmetric  & 2.50 & 2.50 & 1.25 & 3.75  \\ \cline{2-6} 
                                 & Asymmetric & 2.50 & 1.75 & 1.25 & 3.00  \\ \hline
\multirow{2}{*}{Performance}     & Symmetric  & 5.50 & 0.75 & 5.25 & 6.00  \\ \cline{2-6} 
                                 & Asymmetric & 5.50 & 1.75 & 5.00 & 6.75  \\ \hline
\end{tabular}
\caption{NASA-TLX results.}
\label{tab:tlx}
\end{table*}

\begin{table*}[h]
\begin{tabular}{|l|l|l|l|l|l|}
\hline
\textbf{Modality} & \textbf{Participant Type} & \textbf{Median} & \textbf{IQR} & \textbf{Q1} & \textbf{Q3} \\ \hline
\multirow{2}{*}{Symmetric}  & Presenter & 5.75 & 2.03 & 4.22 & 6.25  \\ \cline{2-6} 
                            & Audience  & 6.12 & 0.34 & 6.00 & 6.34  \\ \hline
\multirow{2}{*}{Asymmetric} & Presenter & 5.31 & 0.72 & 5.00 & 5.72  \\ \cline{2-6} 
                            & Audience  & 4.62 & 0.38 & 4.53 & 4.91  \\ \hline
\end{tabular}
\caption{Group awareness questionnaire responses.}
\label{tab:awareness_results}
\end{table*}

\begin{table*}[h]
\begin{tabular}{|l|l|l|l|l|l|}
\hline
\textbf{Modality} & \textbf{Participant Type} & \textbf{Median} & \textbf{IQR} & \textbf{Q1} & \textbf{Q3} \\ \hline
\multirow{2}{*}{Symmetric}  & Presenter & 6.39 & 1.11 & 5.75 & 6.86  \\ \cline{2-6} 
                            & Audience  & 6.50 & 0.53 & 6.36 & 6.89  \\ \hline
\multirow{2}{*}{Asymmetric} & Presenter & 6.06 & 1.25 & 5.47 & 6.72  \\ \cline{2-6} 
                            & Audience  & 5.94 & 0.77 & 5.56 & 6.33  \\ \hline
\end{tabular}
\caption{Engagement questionnaire responses.}
\label{tab:engagement_results}
\end{table*}

\newpage
\subsection{Observational Data}
In these tables, IQR indicates the interquartile range of the data. Q1 and Q3 represent the first quartile (25\%) and the third quartile (75\%) of the data, respectively.

\begin{table*}[h]
\begin{tabular}{|l|l|l|l|l|l|}
\hline
\textbf{Modality} & \textbf{Participant Type} & \textbf{Median} & \textbf{IQR} & \textbf{Q1} & \textbf{Q3} \\ \hline
\multirow{2}{*}{Symmetric}  & Presenter & 0.50 & 2.50 & 0.00 & 2.50    \\ \cline{2-6} 
                            & Audience  & 2.50 & 1.75 & 1.25 & 3.00   \\ \hline
\multirow{2}{*}{Asymmetric} & Presenter & 1.50 & 4.50 & 0.00 & 4.50    \\ \cline{2-6} 
                            & Audience  & 4.50 & 13.50 & 0.00 & 13.50  \\ \hline
\end{tabular}
\caption{Counts of social expressions observed.}
\label{tab:social_expr}
\end{table*}

\begin{table*}[h]
\begin{tabular}{|l|l|l|l|l|l|}
\hline
\textbf{Modality} & \textbf{Participant Type} & \textbf{Median (\%)} & \textbf{IQR (\%)} & \textbf{Q1 (\%)} & \textbf{Q3} \\ \hline
\multirow{2}{*}{Symmetric}  & Presenter & 5.65 & 8.30 & 1.65 & 9.95    \\ \cline{2-6} 
                            & Audience  & 5.45 & 8.20 & 3.80 & 12.00   \\ \hline
\multirow{2}{*}{Asymmetric} & Presenter & 6.65 & 10.05 & 1.15 & 11.20  \\ \cline{2-6} 
                            & Audience  & 2.80 & 4.33 & 1.35 & 5.68    \\ \hline
\end{tabular}
\caption{Percentage of presentation participants looked at each other.}
\label{tab:perc_lookat}
\end{table*}

\begin{table*}[h]
\begin{tabular}{|l|l|l|l|l|}
\hline
\textbf{Modality} & \textbf{Median} & \textbf{IQR} & \textbf{Q1} & \textbf{Q3}  \\ \hline
Symmetric         & 2.00 & 3.50 & 1.00 & 4.50  \\ \hline
Asymmetric        & 2.50 & 3.25 & 0.50 & 3.75  \\ \hline
\end{tabular}
\caption{Counts of occurrences at which participants made eye contact with each other.}
\label{tab:eye_occurrence}
\end{table*}

\begin{table*}[h]
\begin{tabular}{|l|l|l|l|l|}
\hline
\textbf{Modality} & \textbf{Median} & \textbf{IQR} & \textbf{Q1} & \textbf{Q3} \\ \hline
Symmetric         & 0.38  & 2.52 & 0.08 & 2.60  \\ \hline
Asymmetric        & 1.00  & 2.47 & 0.18 & 2.65  \\ \hline
\end{tabular}
\caption{Percentage of the presentation that participants made eye contact with each other.}
\label{tab:eye_percent}
\end{table*}

\begin{table*}[h]
\begin{tabular}{|l|l|l|l|l|l|}
\hline
\textbf{Modality} & \textbf{Label} & \textbf{Median} & \textbf{IQR} & \textbf{Q1} & \textbf{Q3} \\ \hline
\multirow{2}{*}{Symmetric}  & Presenter & 11.50 & 13.75 & 7.25 & 21.00 \\ \cline{2-6} 
                            & Audience  & 5.50  & 1.00 & 5.00  & 6.00  \\ \hline
\multirow{2}{*}{Asymmetric} & Presenter & 8.50  & 6.80 & 4.00  & 10.80 \\ \cline{2-6} 
                            & Audience  & 7.00  & 5.25 & 2.50  & 7.75  \\ \hline
\end{tabular}
\caption{Counts of total references to presentation content participants made.}
\label{tab:ref_count}
\end{table*}

\begin{table*}[]
\begin{tabular}{|l|l|l|l|l|l|}
\hline
\textbf{Reference} & \textbf{Modality} & \textbf{Median} & \textbf{IQR} & \textbf{Q1} & \textbf{Q3} \\ \hline
\multirow{2}{*}{Pointing}                   & Symmetric  & 6.00 & 7.45 & 3.75 & 11.20  \\ \cline{2-6} 
                                            & Asymmetric & 3.50 & 6.25 & 1.50 & 7.75   \\ \hline
\multirow{2}{*}{Deictic Speech}             & Symmetric  & 5.50 & 8.25 & 1.00 & 9.25   \\ \cline{2-6} 
                                            & Asymmetric & 1.50 & 1.00 & 1.00 & 2.00   \\ \hline
\multirow{2}{*}{Relative Language}          & Symmetric  & 0.00 & 0.75 & 0.00 & 0.75  \\ \cline{2-6} 
                                            & Asymmetric & 1.00 & 1.50 & 0.25 & 1.75  \\ \hline
\multirow{2}{*}{Verbal Property References} & Symmetric  & 0.00 & 0.75 & 0.00 & 0.75  \\ \cline{2-6} 
                                            & Asymmetric & 1.00 & 0.50 & 0.25 & 0.75  \\ \hline
\multirow{2}{*}{Verbal Reference Chaining}  & Symmetric  & 0.00 & 0.00 & 0.00 & 0.00  \\ \cline{2-6} 
                                            & Asymmetric & 0.00 & 0.00 & 0.00 & 0.00  \\ \hline
\end{tabular}
\caption{Counts of reference types by presenter modality.}
\label{tab:ref_type}
\end{table*}

\newcommand{\cellwmodality}{1.5cm}
\newcommand{\cellwref}{4cm}

\begin{table*}[]
\begin{tabular}{|p{\cellwmodality}|p{\cellwref}|p{1.75cm}|p{1cm}|p{1cm}|p{1cm}|p{1cm}|}
\hline
\textbf{Modality} & \textbf{Behavior} & \textbf{Participant Type} & \textbf{Median} & \textbf{IQR} & \textbf{Q1} & \textbf{Q3} \\ \hline
\multirow{10}{*}{Asymmetric} & \multirow{2}{*}{Pointing}                  & Audience  & 4.50 & 4.00 & 1.75 & 5.75  \\ \cline{3-7} 
                             &                                            & Presenter & 3.50 & 6.25 & 1.50 & 7.75  \\ \cline{2-7} 
                             & \multirow{2}{*}{Deictic speech}            & Audience  & 1.50 & 3.00 & 0.00 & 3.00  \\ \cline{3-7} 
                             &                                            & Presenter & 1.50 & 1.00 & 1.00 & 2.00  \\ \cline{2-7} 
                             & \multirow{2}{*}{Relative language}         & Audience  & 0.00 & 0.00 & 0.00 & 0.00  \\ \cline{3-7} 
                             &                                            & Presenter & 1.00 & 1.50 & 0.25 & 1.75  \\ \cline{2-7} 
                             & \multirow{2}{*}{Verbal property reference} & Audience  & 0.00 & 0.00 & 0.00 & 0.00  \\ \cline{3-7} 
                             &                                            & Presenter & 0.00 & 3.00 & 0.00 & 3.00  \\ \cline{2-7} 
                             & \multirow{2}{*}{Referential chaining}      & Audience  & 0.00 & 0.00 & 0.00 & 0.00  \\ \cline{3-7} 
                             &                                            & Presenter & 0.00 & 0.00 & 0.00 & 0.00  \\ \hline
\multirow{10}{*}{Symmetric}  & \multirow{2}{*}{Pointing}                  & Audience  & 3.00 & 3.50 & 1.25 & 4.75  \\ \cline{3-7} 
                             &                                            & Presenter & 6.00 & 7.50 & 3.75 & 11.25 \\ \cline{2-7} 
                             & \multirow{2}{*}{Deictic speech}            & Audience  & 1.00 & 1.50 & 0.25 & 1.75  \\ \cline{3-7} 
                             &                                            & Presenter & 5.50 & 8.25 & 1.00 & 9.25  \\ \cline{2-7} 
                             & \multirow{2}{*}{Relative language}         & Audience  & 0.00 & 0.00 & 0.00 & 0.00  \\ \cline{3-7} 
                             &                                            & Presenter & 0.00 & 0.75 & 0.00 & 0.75  \\ \cline{2-7} 
                             & \multirow{2}{*}{Verbal property reference} & Audience  & 0.00 & 0.75 & 0.00 & 0.75  \\ \cline{3-7} 
                             &                                            & Presenter & 0.00 & 0.75 & 0.00 & 0.75  \\ \cline{2-7} 
                             & \multirow{2}{*}{Referential chaining}      & Audience  & 0.00 & 0.00 & 0.00 & 0.00  \\ \cline{3-7} 
                             &                                            & Presenter & 0.00 & 0.00 & 0.00 & 0.00  \\ \hline
\end{tabular}
\caption{Counts of behavior types observed during presentations.}
\label{tab:ref_type}
\end{table*}
\newpage

\section*{Disclaimer}
This paper was prepared for informational purposes by the Global Technology Applied Research center of JPMorganChase. This paper is not a product of the Research Department of JPMorganChase or its affiliates. Neither JPMorganChase nor any of its affiliates makes any explicit or implied representation or warranty and none of them accept any liability in connection with this paper, including, without limitation, with respect to the completeness, accuracy, or reliability of the information contained herein and the potential legal, compliance, tax, or accounting effects thereof. This document is not intended as investment research or investment advice, or as a recommendation, offer, or solicitation for the purchase or sale of any security, financial instrument, financial product or service, or to be used in any way for evaluating the merits of participating in any transaction.

\end{document}